\documentstyle[12pt]{article}
\title{Pauli operator and Aharonov Casher theorem for
measure valued magnetic fields}
\author{L\'aszl\'o Erd\H os
 \thanks{Email address: {\tt lerdos@math.gatech.edu}.
Partially supported by NSF grant DMS-9970323}
\\ School of Mathematics, Georgiatech, Atlanta GA 30332 \\
and \\
Vitali Vougalter  \thanks{Email address: {\tt vitali@math.ubc.ca}}
\\ Department of Mathematics,
 University of British Columbia \\
Vancouver, B.C., Canada V6T 1Z2}

\date{Original: May 3, 2001. Revision: Aug 28, 2001}

\newtheorem{theorem}{Theorem}[section]
\newtheorem{proposition}[theorem]{Proposition}
\newtheorem{corollary}[theorem]{Corollary}
\newtheorem{lemma}[theorem]{Lemma}
\newtheorem{definition}[theorem]{Definition}
\newtheorem{conjecture}[theorem]{Conjecture}
\newtheorem{counterexample}[theorem]{Counterexample}

\newcommand{\rd}{{\rm d}}
\newcommand{\be}{\begin{equation}}
\newcommand{\ee}{\end{equation}}
\newcommand{\bey}{\begin{eqnarray}}
\newcommand{\eey}{\end{eqnarray}}

\newcommand{\sfrac}[2]{{\textstyle \frac{#1}{#2}}}

\newcommand{\triple}{ |\! |\! |}

\newcommand{\bA}{{\bf A}}
\newcommand{\bZ}{{\bf Z}}

\newcommand{\bsigma}{\mbox{\boldmath $\sigma$}}

\newcommand{\cp}{{\pi}}

\newcommand{\bR}{{\bf R}}
\newcommand{\bC}{{\bf C}}

\newcommand{\ep}{\varepsilon}
\newcommand{\wh}{\widehat}
\newcommand{\wt}{\widetilde}
\newcommand{\Av}{\mbox{Av}}

\newcommand{\cC}{{\cal C}}

\newcommand{\cP}{{\cal P}}
\newcommand{\cD}{{\cal D}}

\newcommand{\cK}{{\cal K}}

\newcommand{\cM}{{\cal M}}

\begin{document}
\maketitle

\begin{abstract}
We define the two dimensional Pauli operator 
and identify its core for 
magnetic fields that are regular Borel measures. The magnetic
field is generated by a scalar potential hence we bypass the usual
$\bA\in L^2_{loc}$ condition on the vector potential which
does not allow to consider such singular fields. 
We extend the  Aharonov-Casher theorem for magnetic fields that are
measures with finite total variation and we present a counterexample
in case of infinite total variation. One
of the key technical tools is a weighted $L^2$ estimate on
a singular integral operator.
\end{abstract}


\medskip\noindent
{\it Running title:} Pauli Operator for  Measure Valued Fields.

\section{Introduction}

We consider the usual Pauli operator in $d=2$ dimensions with a magnetic
field $B$
$$
	 H = \big[\bsigma \cdot (-i\nabla + \bA) \big]^2
	=  (-i\nabla + \bA)^2 + \sigma_3B \qquad \mbox{on}
	\quad L^2(\bR^2, \bC^2),
$$
$B:=\mbox{curl}(\bA) = \nabla^\perp\!\cdot\! \bA$
with $\nabla^\perp : = (-\partial_2, \partial_1)$.
Here $\bsigma \cdot (-i\nabla + \bA)$
is the two dimensional Dirac operator on the trivial spinorbundle
over $\bR^2$ with real vector potential $\bA$
and $\bsigma = (\sigma_1, \sigma_2)$ are the first two
 Pauli matrices.  Precise conditions on $\bA$ and $B$
will be specified later.

The Aharonov-Casher Theorem \cite{AC} states that
the dimension of the kernel of $H$ is given
\be
	\mbox{dim}\; \mbox{Ker} (H) 
 =\lfloor |\Phi|\rfloor,
\label{ACeq}\ee
where
$$
	\Phi:= \frac{1}{2\pi}\int_{\bR^2} B(x) \rd x
$$
(possibly $\pm \infty$)
is the flux (divided by $2\pi$)
 and $\lfloor \quad \rfloor$ denotes the lower integer
part ($\lfloor n \rfloor = n-1$ for $n\ge 1$ integer and $\lfloor 0\rfloor=0$).
Moreover, $\sigma_3 \psi= -s\psi$ for any $\psi\in \mbox{Ker} (H)$,
where $s=\mbox{sign}(\Phi)$.

On a $Spin^c$-bundle over
 $S^2$ with a smooth magnetic field the analogous theorem 
is equivalent to the index theorem (for a short direct proof 
see \cite{ES}).  From topological reasons the analogue of $\Phi$,
 the total curvature of a
connection,
is an integer (the Chern number of the determinant line bundle),
 and the number of zero modes
of the corresponding Dirac operator  is $\big| \Phi\big|$.

\medskip

In the present paper we investigate two related questions:

\medskip

(i) What is the most general class of magnetic fields for which
the Pauli operator can be properly defined on $\bR^2$?

\medskip

(ii) What is the most general class of magnetic fields
 for the Aharonov-Casher theorem to hold on $\bR^2$?

\medskip

Pauli operators are usually defined either via the magnetic
Schr\"odinger operator, $(-i\nabla +\bA)^2$, by adding
the magnetic field $\sigma_3 B$ as an external potential,
or directly by the  quadratic form of the Dirac operator
$\bsigma \cdot (-i\nabla + \bA)$  (see Section \ref{stand}).
In both ways, the standard condition  $\bA\in L^2_{loc}(\bR^2, \bR^2)$ is
necessary. 

On the other hand, the statement of the Aharonov-Casher theorem
uses only that $B\in L^1(\bR^2)$, and in fact
$B$ can even be a measure.
It is therefore a natural question to extend the Pauli operator
for such magnetic fields and investigate the validity of
the Aharonov-Casher theorem.
 However, even if $B\in L^1$, it
might not be generated by an $\bA\in L^2_{loc}$. For example,
 any gauge $\bA$ generating 
the radial field $B(x)= |x|^{-2}| \log |x| \; |^{-3/2} {\bf 1}(|x|\leq
\sfrac{1}{2}) \in L^1$
satisfies $\int_{|x|\leq 1/2} |\bA(x)|^2 \rd x \ge \int_0^{1/2} 
(r|\log r|)^{-1} \rd r =\infty$ (here ${\bf 1}$ is the characteristic
function).
Hence the Pauli operator cannot be defined in the usual
way on  $C_0^\infty$ as its core. In case of a point singularity
at $p\in\bR^2$ one can study the extensions from $C_0^{\infty}(\bR^2\setminus
\{ p\})$, but such approach may not be possible
 for $B$ with a more complicated
singular set.

In this paper we present an alternative method which enables
us to define the Pauli operator for any magnetic field that is
a regular Borel measure (Theorem \ref{bm}).
Moreover, we actually define the corresponding quadratic form
on the maximal domain and identify a core.
We recall that the maximal domain contains all finite energy states, hence
it has a direct physical interpretation. For mathematical analysis, however,
one  needs to know a core explicitly that contains reasonably ``nice''
functions.
For most Schr\"odinger type operators the core consists of smooth
functions. In case of Pauli operators with singular magnetic fields
 the core will be identified as the set of
smooth functions times an explicit nonsmooth factor.

\bigskip

The basic idea is to define the Pauli operator via a real generating potential
function $h$, satisfying 
\be
	\Delta h = B
\label{poisson}\ee
 instead of the
usual vector potential $\bA$. This
potential function appears in the original proof of the
Aharonov-Casher theorem.
The key identity is the following
\be
	\int\Big|\bsigma \cdot (-i\nabla + \bA) \psi\Big|^2
	= 4 \int \Big|\partial_{\bar z} (e^{-h}\psi_+)\Big|^2e^{2h}
	+ \Big|\partial_{z} (e^{h}\psi_-)\Big|^2e^{-2h}
\label{quad}\ee
for regular data, with
 $\bA := \nabla^\perp h$ (integrals without specified domains
are understood on $\bR^2$ with respect to the Lebesgue measure).
 We will {\it define} the Pauli quadratic form
by the right hand side even for less regular data.
It turns out that any magnetic field that is a regular Borel measure can be 
handled by an $h$-potential.

The main technical tool is that for an appropriate choice of $h$,
the weight function $e^{\pm 2h}$ (locally) belongs to
the Muckenhoupt $A_2$ class (\cite{GR}, \cite{St}). Therefore the
maximal operator and certain singular integral operators
are bounded on the weighted $L^2$ spaces. This will
be essential to identify the core of the Pauli operator.

We point out that this approach does {\it not} apply to the
magnetic Schr\"odinger operator $(-i\nabla +\bA)^2$.

\bigskip

The Aharonov-Casher theorem has been rigorously proven
only for a restricted class of magnetic fields on $\bR^2$.
The conditions involve some control on the decay 
at infinity  and on local singularities.
 In fact, to our knowledge,
the optimal conditions have never been investigated.
The original paper \cite{AC} does not focus on 
 conditions. The exposition \cite{CFKS} assumes
 compactly supported bounded magnetic
field $B(x)$. The Ph.D. thesis by K. Miller \cite{Mi} assumes boundedness,
and assumes that
$\int |B(x)| \; \Big| \log |x|\; \Big|  \rd x < \infty$.
The boundedness condition is clearly too strong, and
it can be easily replaced
with the assumption that $B\in \cK(\bR^2)$ Kato class.

Miller also observes that in case of integer $\Phi\neq0$
there could be either $|\Phi|$ or $|\Phi|-1$ zero states,
but if the field is compactly supported then the number
of states is always $|\Phi|-1$ \cite{CFKS}.

\medskip

The idea behind each proof is
to construct a potential function $h$ satisfying (\ref{poisson}).
{\it Locally}, $H\psi =0$ is equivalent to $\psi = (e^hg_+, e^{-h}g_-)$
with $\partial_{\bar z}g_+=0$, $\partial_{ z}g_-=0$, where
we identify $\bR^2$ with $\bC$ and use the
notations $x =(x_1, x_2)\in \bR^2$ and $z=x_1+ix_2 \in \bC$
simultaneously.
 The condition $\psi\in L^2(\bR^2,\bC^2)$ together with the explicit
growth (or decay) 
rate of $h$ at infinity determines the {\it global} solution space
by identifying the space of (anti)holomorphic functions $g_\pm$
with a controlled growth rate at infinity.

For bounded magnetic fields
decaying fast enough at infinity,
a solution to (\ref{poisson}) is given by
\be
	h(x) = {1\over 2\pi} \int_{\bR^2} \log |x-y| B(y) \rd y 
\label{trivsol}
\ee
and $h(x)$ behaves as $ \approx\Phi \log |x|$ for
large $x$. If $\Phi\ge 0$, then
$e^hg_+$ is never in $L^2$, and $e^{-h}g_-\in L^2$
if  $|g_-|$  grows at most as the
 $\Big(\lfloor\Phi\rfloor -1\Big)$-th power of $|x|$.
If $\Phi <\infty$, then $g_-$ must be a polynomial
of degree at most $\lfloor\Phi\rfloor -1$.
If $\Phi= \infty$, then the integral in (\ref{trivsol}) 
is not absolutely convergent. If the radial behavior of $B$
is regular enough, then $h$ may still be defined via (\ref{trivsol})
as a conditionally convergent integral and we then
have a solution space of infinite dimension.

\medskip

Conditions on local regularity and decay at infinity are
used to establish bounds on the auxiliary function 
$h$ given by (\ref{trivsol}), but they are not a priori
needed for   the Aharonov-Casher
 Theorem (\ref{ACeq}). We show that local regularity conditions
are irrelevant by proving the Aharonov-Casher
theorem for any measure valued magnetic fields with finite
total variation (Theorem \ref{ACprec}). Many fields with
infinite total variation can also be covered; some
regular behavior at infinity is sufficient (Corollary \ref{reg}).
However, some control is needed in general, as we
present a counterexample
to the Aharonov-Casher theorem for a magnetic field
with infinite total variation.
\begin{counterexample}\label{cont}
There exists a continuous bounded magnetic field $B$ such that
$\int_{\bR^2} |B| =\infty$ and
\be
	\Phi := \lim_{r\to\infty} \Phi(r)
	= \lim_{r\to\infty} {1\over 2\pi} \int_{|x|\leq r} B(x) \rd x
\label{impr}\ee
exists and $\Phi >1$,
but $\mbox{dim}\; \mbox{Ker} H = 0$.
\end{counterexample}

\medskip

Finally, we recall a conjecture from \cite{Mi}:

\begin{conjecture}\label{millconj1}
Let $B(x)\ge 0$ with flux
$\Phi := \sfrac{1}{2\pi}\int B$, which may be infinite.
Then the dimension of $\mbox{Ker}(H)$ is at least
$\lfloor \Phi\rfloor$.
\end{conjecture}

The proof in \cite{Mi} failed because it would have relied on the 
conjecture that for  any continuous function $B\ge 0$
there exists a positive
solution $h$ to (\ref{poisson}). This is false.
A counterexample (even with finite $\Phi$)
was given by C. Fefferman and B. Simon and it
was presented in \cite{Mi}. However, the same magnetic field 
does {\it not} yield a counterexample to Conjecture \ref{millconj1}.

Theorem \ref{ACprec} settles this conjecture for $\Phi<\infty$,
but the case  $\Phi =\infty$ remains open. The magnetic field
in our counterexample does not have a definite sign, in fact
 $\Phi$ is defined only as an improper integral.

\section{Definition of the Pauli operator}\label{sec:def}

\subsection{Standard definition for $\bA\in L^2_{loc}$}\label{stand}

The standard definition
of the {\it magnetic Schr\"odinger operator},
 $(-i\nabla +\bA)^2$,  or the {\it Pauli operator},
$[\bsigma\cdot(-i\nabla +\bA)]^2$, as a quadratic
form, requires $\bA \in L^2_{loc}$ 
(see e.g. \cite{LS, LL} and  for the Pauli operator
\cite{Sob}). We define
$\Pi_k := -i\partial_k +A_k$, $Q_\pm: = \Pi_1 \pm i\Pi_2$,
or with complex notation 
$Q_+ = -2i\partial_{\bar z} + a$,
$Q_- = -2i\partial_{z} + \bar a$
with $a: = A_1 + iA_2$.
These are closable operators, originally defined on $C_0^\infty(\bR^2)$.
Their closures are denoted by the same letter on the minimal
domains $\cD_{min}(\Pi_j)$ and $\cD_{min}(Q_\pm)$.

Let
$$
	s_\bA(u,u) : = \| \Pi_1 u\|^2 +  \| \Pi_2 u\|^2 
	= \int | (-i\nabla +\bA)u|^2 \; ,\qquad 
	u\in C_0^\infty(\bR^2)
$$
be the closable quadratic form associated with the
magnetic Schr\"odinger operator on the minimal form 
domain $\cD_{min} (s_\bA)$. It is known \cite{Si} that the minimal domain
 coincides with the maximal domain $\cD_{max}(s_\bA):=\{
u\in L^2(\bR^2)\; : \; s_\bA(u,u)<\infty\}$. We will denote $\cD(s_\bA):=
\cD_{max}(s_\bA)= \cD_{min}(s_\bA)$ and let $S_\bA$ be the corresponding
self-adjoint operator.

The closable quadratic form associated with the {\it Pauli operator} is
$$
	p_\bA(\psi, \psi): =  \| Q_+\psi_+\|^2+
	\| Q_-\psi_-\|^2 =
	\int | \bsigma\cdot(-i\nabla +\bA)\psi|^2 \; , 
	\qquad \psi = \pmatrix{\psi_+\cr \psi_-}
	\in C_0^\infty(\bR^2, \bC^2) \; .
$$
The condition $\bA\in L^2_{loc}$ is obviously necessary.
The minimal form domain is $\cD_{min} (p_\bA) 
= \cD_{min}(Q_+)\otimes\cD_{min}(Q_-)$, while $\cD_{max}(p_\bA)
:= \{ \psi \in L^2(\bR^2, \bC^2)\; : \; p_\bA(\psi, \psi)<\infty\}$.
The unique self-adjoint operators associated with
these forms are $P_\bA^{min}$ and $P_\bA^{max}$.

Clearly $\cD_{min} (p_\bA)\subset \cD_{max} (p_\bA)$.
 For a locally bounded  magnetic field 
$B=\nabla^\perp\!\cdot\! \bA$ one can choose
a vector potential $\bA\in L^\infty_{loc}$ by the Poincar\'e formula and
in this case $\cD_{min} (p_\bA)=\cD_{max} (p_\bA)$, i.e., $P_\bA^{min}
=P_\bA^{max}$. To see
this, we first approximate any $\psi\in \cD_{max} (p_\bA)$
in the norm $[\| \cdot \|^2 + p_\bA(\cdot, \cdot)]^{1/2}$
by functions $\psi_n=\psi\chi_n$ of compact support, where $\chi_n\to 1$
and $\|\nabla \chi_n\|_\infty\to0$. Then we use that
$ \|\nabla\psi_n\|^2\leq 2p_\bA(\psi_n, \psi_n)+
2\|\bA \psi_n\|^2<\infty$, i.e. $\psi_n \in H^1$,
so it can be appoximated by $C_0^\infty$ functions in $H^1$
and also in $[\| \cdot \|^2 + p_\bA(\cdot, \cdot)]^{1/2}$.

To our knowledge, the precise conditions  for 
$\cD_{min} (p_\bA)=\cD_{max} (p_\bA)$ have not been investigated
in general. Such a result is
expected to be harder than $\cD_{min} (s_\bA)=\cD_{max} (s_\bA)$
due to the lack of the diamagnetic inequality.
In the present paper we do not address this question.
We will define the Pauli quadratic form differently 
and always on the appropriate maximal domain
since this is the physically relevant object (finite energy)
and we identify a natural core for computations.
We will see that this approach works for data even more singular
than $\bA\in L^2_{loc}$ and for $\bA\in L^2_{loc}$ we obtain $P_\bA^{max}$
back.  It is nevertheless a mathematically
interesting open question to determine 
the biggest subset of $L^2_{loc}$ vector potentials such that
the set $C_0^\infty$ is still a core for the Pauli form.

Finally we remark   $\cD (s_\bA)\otimes\cD (s_\bA)
  \subset \cD_{min} (p_\bA)$ for $\bA\in L^2_{loc}$.
 In case of $B=\nabla^\perp\!\cdot\! \bA \in L^\infty$, these
two domains are equal and 
$P_\bA^{min} = S_\bA\otimes I_2 + \sigma_3 B$.
If $B\in L^\infty_{loc}$ only, then the form domains coincide
locally. For more details on these statements, see Section 2 of \cite{Sob}.

\subsection{Measures and integer point fluxes}

Let $\cM$ be the set of signed real 
Borel measures $\mu(\rd x)$ on $\bR^2$ with finite
total variation, $|\mu|(\bR^2)=\int_{\bR^2} |\mu|(\rd x)<\infty$.
Let $\overline\cM$ be the
set of signed real regular  Borel measures $\mu$ on $\bR^2$,
in particular they have
$\sigma$-finite total variation.
If $\mu(\rd x) = B(x)\rd x$ is absolutely continuous, then $\mu\in\cM$ is
equivalent to $B\in L^1$.
Let  $\overline{\cM^*}$
 be the set of all measures $\mu\in \overline{\cM}$ such that
$\mu(\{x \})\in (-2\pi,2\pi)$ for any point $x\in \bR^2$,
and  $\cM^*: = \cM \cap \overline{\cM^*}$.

\begin{definition}\label{mustdef}
Two measures $\mu, \mu'\in \overline{\cM}$
 are said to be {\bf equivalent} if $\mu-\mu' = 2\pi\sum_j n_j \delta_{z_j}$,
where $n_j\in \bZ$, $z_j\in\bR^2$.
The equivalence class of any measure $\mu\in  \overline{\cM}$ contains
a unique measure, called the {\bf reduction} of $\mu$
and denoted by $\mu^*$,
 such that $\mu^*(\{x\})\in [-\pi, \pi)$ for any $x\in \bR^2$.
In particular, $\mu^*\in \overline{\cM^*}$. 
\end{definition}

The Pauli operator associated with  $\mu\in \overline{\cM}$
will depend only on the equivalence class of $\mu$
up to a gauge transformation, so we can work with
  $\mu\in \overline{\cM^*}$.
This just reflects the  physical expectation 
that any magnetic point flux $2\pi n\delta_z$, with integer
$n$, is removable by the gauge transformation
$\psi(x) \to e^{in \varphi}\psi(x)$, where $\varphi = \mbox{arg}(x-z)$.
In case of several point fluxes, $2\pi\sum_j n_j \delta_{z_j}$,
the phase factor should be $\exp\Big( i\sum_j n_j \mbox{arg}(x-z_j)\Big)$,
but it may not converge for an infinite set of points $\{ z_j \}$.

However, any $\mu\in \overline{\cM}$
can be uniquely written as $\mu = \mu^* + 2\pi\sum_j n_j \delta_{z_j}$
with $n_j\in {\bf Z}\setminus \{ 0 \}$ and
 with a set of distinct points $\{ z_j\}$
which do not accumulate in $\bR^2 \equiv \bC$. Let $I_+ : = \{ j \; : \;
n_j >0\}$, and  $I_- : = \{ j \; : \; n_j <0\}$ be the set of indices of
the points with positive and negative masses, respectively.
By Weierstrass theorem, there exist
analytic functions $F_\mu(x)$ and $G_\mu(x)$  (recall $x=x_1+ix_2$)
such that $F_\mu$  has zeros exactly at the
 points $\{ z_j \; : \; j\in I_+\}$
with multiplicities $n_j$,  and
$G_{\mu}$ has zeros at $\{ z_{j}\; : \; j\in I_-\}$
 with multiplicities $-n_{j}$.
 Let $L_\mu(x): =F_\mu(x) \overline{G}_{\mu}(x)$.
 Then the
integer point fluxes can be removed by the unitary 
gauge transformation
\be
	U_\mu: \psi(x) \to {L_\mu(x)\over |L_\mu(x)|} 
	\, \psi(x)\; .
\label{umudef}\ee
For example, for any compact set $K\subset \bR^2$, we can write
 $L_\mu/|L_\mu|$ as
$$
L_\mu(x)/|L_\mu(x)| = \exp\Big( i \sum_{j\; : \; z_j\in K}
n_j \mbox{arg}(x-z_j) + i H_K(x)\Big)\, , \qquad x\in K\, ,
$$
where $H_K$ is a real harmonic function on $K$.
In particular, for any $\psi$ supported on $K$,
$$
U_\mu^* (-i\nabla)U_\mu \psi
= \Big(-i\nabla + \sum_{j\; : \; z_j\in K} n_j \bA_j + \nabla H_K\Big)\psi\; ,
$$
where $\nabla^\perp\cdot \bA_j = 2\pi \delta_{z_j}$.

\subsection{Potential function}\label{sec:pot}

The Pauli quadratic form for magnetic fields $\mu\in \overline{\cM^*}$
 will be defined via the right hand side of
 (\ref{quad}), where $h$ is a solution to $\Delta h =\mu$.
The following theorem shows that for $\mu\in \cM^*$ one can
always choose a good potential function $h$.
Later we will extend it for $\mu\in \overline{\cM^*}$.

\begin{theorem}\label{thm:hdef}
 Let $\mu\in \cM^*$ and $\Phi: = {1\over 2\pi}\int \mu(\rd x)$
 be the total flux (divided by $2\pi$). There is  $0<\ep(\mu)\leq 1$
such that for any
$0< \ep < \ep(\mu)$  there exists a real valued function $h=h^{(\ep)}
\in \cap_{p<2}W^{1,p}_{loc}$ with $\Delta h= \mu$
(in distributional sense), such that

(i) For any compact set $K\subset \bR^2$ and any
square $Q\subset K$
\be
	\Big( {1\over |Q|}\int_Q e^{2h}\Big)
	\Big( {1\over |Q|}\int_Q e^{-2h}\Big) \leq C_1(K,\ep,\mu)
\label{A2}\ee

(ii) $e^{\pm 2h}\in L_{loc}^{1+\ep}$.

(iii)  $h$ can be split as 
$h = h_1 + h_2$ with the following estimates:
\be
	\Bigg| \; {  h_1(x)\over \log |x|} - \Phi \; \Bigg|\leq \ep
	\qquad \mbox{for} \quad |x|\ge R(\ep,\mu)
\label{mainh}\ee
and
\be
	\int_{Q(u)} e^{\pm 2 h_2} \leq C_2(\ep) \langle u\rangle^{2\ep}
\label{errorh}
\ee
with some  constants $C_1(K,\ep,\mu), C_2(\ep)$ and $R(\ep,\mu)$.
Here $Q(u)= [u-\sfrac{1}{2}, u+\sfrac{1}{2}]^2$ 
denotes the unit square about 
$u\in\bR^2$ and $\langle u\rangle = (u^2+1)^{1/2}$.
\end{theorem}

{\it Remark.} The property (i) means that $e^{2h}$ satisfies
a certain   reversed H\"older inequality locally. 
 If  (\ref{A2}) 
were true for any square $Q\subset \bR^2$ with a $K$-independent
constant, then $e^{2h}$ would be in
the weight-class $A_2$ used in harmonic analysis (see \cite{GR,St}). 
Nevertheless, this property will allow us to use
weighted $L^2$-bounds on a certain singular integral operator
locally (Lemma \ref{local}).

We also remark  that property (ii) follows from the local analog
of the well-known fact that  $\omega \in A_2 \Longrightarrow \omega 
\in A_{p}$ for some $p<2$.

\begin{corollary}\label{w11}
If $h\in L^1_{loc}$ satisfies $\Delta h \in \overline{\cM^*}$, then
$h\in  W^{1,p}_{loc}$ for all $p<2$ and $e^{\pm 2h}\in L^{1}_{loc}$.
If, in addition, $\Delta h \in \cM^*$, then $e^{\pm 2h}\in L^{1+\ep}_{loc}$
with some $\ep>0$. 
\end{corollary}

{\it Proof.} Suppose first that $\mu: = \Delta h
\in  \cM^*$.  Choose $\ep < \ep(\mu)$
and consider $h^{(\ep)} \in \bigcap_{p<2}W^{1,p}_{loc}$ 
 constructed in Theorem \ref{thm:hdef} with 
$e^{\pm 2h^{(\ep)}} \in L^{1+\ep}_{loc}$.
Since $\Delta (h-h^{(\ep)})=0$, we have
$h=h^{(\ep)} + \varphi$ with a smooth function $\varphi$
so the statements follow for $h$ as well.

If $\mu = \Delta h$ has infinite total variation, then
Theorem \ref{thm:hdef} cannot be applied directly.
But for any compact set $K$ one can find another compact
set $K^*$ with $K\subset \mbox{int}( K^*)$ and then
the measure $\Delta h \in \overline{\cM^*}$ restricted to $K^*$
has finite total variation. Therefore  one can find 
a function $h^*\in  \bigcap_{p<2}W^{1,p}_{loc}$,
$e^{\pm h^*}\in L^2_{loc}$ with $\Delta  h^* =\mu$ on
$K^*$, i.e., $h-h^*$ is harmonic on $K^*$, hence it is smooth and
bounded on $K$. So $h\in \bigcap_{p<2} W^{1,p}(K)$
 and $e^{\pm h} \in L^{2}(K)$ follows from the same properties
of $h^*$. $\,\,\Box$

\bigskip

{\it Proof of Theorem \ref{thm:hdef}.} Step 1.
First we write 
$\mu = \mu_d + \mu_c$,
where $\mu_d := 2\pi \sum_j C_j\delta_{z_j}$
 ($C_j\in (-1,1)$, $z_j\in \bC \equiv \bR^2$)
 is the discrete part of the measure $\mu$, and $\mu_c$ is continuous,
i.e., $\mu_c(\{ x \})=0$ for any point $x\in \bR^2$.  
The summation can be infinite, finite or empty, but
$\sum_j |C_j|<\infty$. We also assume that $z_j$'s are distinct.
Let
\be
	\ep(\mu):= {1\over 10}\min_j \Big\{ 1 - |C_j| \Big\}
\label{epdef}\ee
then clearly $\ep(\mu)>0$.
 We fix an $0<\ep <\ep(\mu)$. All objects defined
below will depend on $\ep$, but we will neglect this fact in
the notations.

We split the
measure $\mu_d = \mu_{d,1} +  \mu_{d,2}$ 
such that 
$$
	 \mu_{d,1}
	 := 2\pi \sum_{j=1}^N  C_j\delta_{z_j}  \; ,
	\qquad \mu_{d,2}
	 := 2\pi\sum_{j=N+1}^\infty  C_j\delta_{z_j}  \; ,
$$
where $N$ is chosen such that
$2\pi\sum_{j=N+1}^\infty |C_j| < \ep/2$. 
In particular $|\mu_{d, 2}|(\bR^2)< \ep/2$.

We define
\be
	h_{d,j}(x) : = {1\over 2\pi}\int 
	\log {|x-y|\over \langle y\rangle} \mu_{d,j}(\rd y)\; ,
	 \qquad j=1,2 \; ,
\label{hddef}\ee
so that   $\Delta h_{d,j} = \mu_{d,j}$.
Notice that $ h_{d,j}(x)$ is well defined  for a.e.  $x$,
moreover $ h_{d,j} \in W^{1,p}_{loc}$ for all $p<2$ by Jensen's
inequality.

\bigskip

Step 2. We split $\mu_c = \mu_{c,1}+ \mu_{c,2}$
such that $\mu_{c,1}$ be compactly supported and
$|\mu_{c, 2}|(\bR^2) < \ep/2$. 
We set $\mu_j:= \mu_{d, j}+ \mu_{c,j}$, $j=1,2$.
Then we define
\be
	h_{c,j}(x): = {1\over 2\pi} \int_{\bR^2} \log
	 {|x-y|\over \langle y\rangle} \mu_{c,j}(\rd y)\; , \qquad j=1,2 \; ,
\label{hcdef}\ee
clearly  $h_{c,j} \in W^{1,p}_{loc}$ for all $p<2$, and 
 $\Delta h_{c,j}= \mu_{c,j}$ (in distributional sense).

Finally, we define
\be
	h_1 :=  h_{d,1}+h_{c,1}\; ,\qquad
	 h_2: = h_{d,2}+h_{c,2} \; ,\qquad h: =  h_1 + h_2
\label{htotdef}\ee
and clearly $\Delta h_j = \mu_j$.
Since  $\mu_{d, 1}$ and $\mu_{c,1}$ are compactly supported,
the estimate (\ref{mainh}) is straightforward.
We will also need the notation $\nu: =\mu_{d,2} + \mu_c = \mu_{c,1}+\mu_2$.

\bigskip

Step 3. For any integer $L$
we define $\Lambda_L:= (2^{-L}\bZ)^2 +
(2^{-L-1}, 2^{-L-1})$ to be  the shifted and rescaled integer lattice.
We define the {\it dyadic squares of scale $L$} to be the squares
$$
	 D^{(L)}_{k}: =\Big[ k_1- 2^{-L-1}, k_1+ 2^{-L-1}\Big)
	\times \Big[k_2- 2^{-L-1}, k_2+ 2^{-L-1}\Big)
$$
of side-length $2^{-L}$ about the lattice
points $k=(k_1, k_2)\in \Lambda_L$. 
The squares 
$$
	\wt D^{(L)}_{k} : =\Big[ k_1- 2^{-L}, k_1+ 2^{-L}\Big)
	\times \Big[k_2- 2^{-L}, k_2+ 2^{-L}\Big)
$$
of double side-length with the same center $k$
are called {\it doubled dyadic squares of scale $L$}.
Similarly, the squares 
$$
	\wh D^{(L)}_{k} : =\Big[ k_1- 3\cdot 2^{-L-1}, k_1+ 3\cdot 
	2^{-L-1}\Big)
	\times \Big[k_2- 3\cdot 2^{-L-1}, k_2+ 3\cdot 2^{-L-1}\Big)
$$
are called {\it tripled dyadic squares of scale $L$}.
For a fixed scale $L$
the collection of dyadic squares is denoted by $\cD_L$.
$\wt \cD_L$ and $\wh\cD_L$ denote the set of doubled and tripled
dyadic squares, respectively.
The elements of $\cD_L$ partition $\bR^2$ for each $L$.
Notice also that every square $Q\subset \bR^2$ can be covered by
a doubled dyadic square of area not bigger than a universal
constant times $|Q|$.

\begin{lemma}\label{Mlemma}
There exists  $1\leq M=M(\mu,\ep)<\infty$ such that
$|\mu|(Q) < 2\pi(1-\ep)$ for any $Q\in\wh\cD_M$.
\end{lemma}

{\it Proof.}  We first
 notice that the support of $\mu_{d,1}$ consists of
finitely many points,
hence for large enough $L$ each element of $\wh\cD_L$ contains at most one
point from this support.

Second, since the measure $|\nu| = |\mu_c| + |\mu_{d,2}|$
 does not charge more than $\ep/2$
to any point, we claim that
 there exists a positive integer $1\leq M=M(\mu,\ep)<\infty$ such
that $|\nu|(D) < \ep$ for any dyadic square of scale $M$. We
can choose $M(\mu,\ep)\ge L$.

This statement is clear by a dyadic decomposition; we start with
the partition of $\bR^2$ into dyadic squares of scale $L$. 
There are just finitely many squares $D\in \cD_L$ such that
$|\nu|(D)\ge \ep$. We split
these squares further into four identical dyadic squares.
 If this process stops after
finitely many steps, then we have reached our $M$
as the scale of the finest decomposition. Now
suppose on the contrary that this process never stops.
Then we could find a strictly decreasing
sequence of nested dyadic squares
$D_1 \supset D_2 \supset \ldots$ such that $|\nu|(D_j)\ge \ep$,
but  $|\nu|$
would charge at least $\ep$ weight to their intersection  which is a 
point.

Finally, since $|\mu|= |\mu_{d,1}| + |\nu|$ and every
tripled square  can be covered by 9 dyadic
squares of the same scale, we have
$|\mu|(Q) \leq 9\ep + 2\pi\max_j |C_j| < 2\pi(1-\ep)$
for each $Q\in \wh \cD_M$ for large enough $M$. $\,\,\,\Box$.

\bigskip

Step 4. Now we turn to the proof of (\ref{A2}) and
first we prove it for any doubled dyadic square of big scale.
Let  $Q = \wt D^{(K)}_k\in \wt\cD_K$ be a doubled dyadic square
with $K\ge M$ and let $\wh Q = \wh D^{(K)}_k$ be the corresponding tripled
square with the same center $k\in \Lambda_K$.
We split the measure $\mu$ as
$$
	\mu = \mu^{int} + \mu^{ext} : = {\bf 1}_{\wh Q} \mu +
	{\bf 1}_{\wh Q^c} \mu
$$
with $|\mu| = |\mu^{int}| + |\mu^{ext}|$ and
$h$ is decomposed accordingly as $h = h^{int}+ h^{ext}$ with
$$
	h^{\#}(x) := {1\over 2\pi} \int_{\bR^2} \log
	{|x-y|\over \langle y \rangle}\mu^\# (\rd y) \; ,
$$
where $\# = \mbox{int, ext}$. We also define
$$
	\wt h^{int}(x) := {1\over 2\pi} \int_{\bR^2} \log
	|x-y|\mu^{int} (\rd y) = h^{int}(x) +  
	{1\over 2\pi} \int_{\bR^2} \log \langle y \rangle\mu^{int} (\rd y)\; .
$$

Let $\Av_{Q}h^{ext} : = |Q|^{-1}\int_{Q} h^{ext}$
be the average of $ h^{ext}$ on $Q$. A simple calculation shows
that
\be
	\Big| h^{ext}(x) -\Av_{Q} h^{ext}\Big|
	\leq C |\mu|(\bR^2)\; , \qquad \forall x\in Q
\label{av}
\ee
with a universal constant $C$
using that $\mu^{ext}$ is supported outside of the tripled square.
Therefore
$$
	\Big( {1\over | Q|}\int_{Q} e^{2h}\Big)
	\Big( {1\over | Q|}\int_{ Q} e^{-2h}\Big)
	\leq e^{4C|\mu|(\bR^2)}\Big( {1\over |Q|}\int_{ Q}
	 e^{2\wt h^{int}}\Big)
	\Big( {1\over | Q|}\int_{Q} e^{-2\wt h^{int}}\Big)
$$

We split $\mu^{int}$ into its positive and negative
parts: $\mu^{int} = \mu^{int}_+ - \mu^{int}_-$, we
let  $\phi_\pm: = {1\over 2\pi} \int_{\wh Q} \mu^{int}_\pm\ge 0$.
By Lemma \ref{Mlemma} and $K\ge M$ we have $\phi:=\phi_++\phi_- <(1-\ep)$.

Now we apply Jensen's inequality for the probability
measures $(2\pi\phi_\pm)^{-1} \mu^{int}_\pm$ (if $\phi_\pm\neq0$):
$$
	\int_{ Q} e^{2\wt h^{int}}
	= \int_{ Q} \exp\Big( {1\over2\pi\phi_+}
	 \int_{\wh Q}\log |x-y|^{2\phi_+} \mu^{int}_+(\rd y)\Big)
	 \exp\Big( {1\over2\pi\phi_-}
	 \int_{\wh Q}\log |x-y'|^{-2\phi_-} \mu^{int}_-(\rd y')\Big)
	\rd x
$$
\be
	\leq  \int_{Q} 
	\rd x {1\over 2\pi\phi_+} \int_{\wh Q} \mu^{int}_+(\rd y)
	 {1\over 2\pi\phi_-}\int_{\wh Q} \mu^{int}_-(\rd y')
	|x-y|^{2\phi_+}|x-y'|^{-2\phi_-}
	\leq C(\ep)|Q|^{1+\phi_+-\phi_-}
\label{one}\ee
with an $\ep$-dependent constant. When performing the $\rd x$
integration, we used the fact that $\phi_- < 1-\ep$, hence
the singularity is integrable. 
Similarly, we have 
$$
	\int_{Q} e^{-2\wt h^{int}} \leq C(\ep)| Q|^{1-\phi_++\phi_-}
$$
which completes the proof of (\ref{A2}) for doubled dyadic
squares of scale at least $M$ with a $K$-independent constant.

\bigskip

Step 5. Next, we prove $e^{\pm 2h}\in L_{loc}^{1+\ep}$. We can follow
the argument in Step 4. On any square $Q\in \wt\cD_M$ we can 
use that $h^{ext}$ is bounded by (\ref{av}) and we can
focus on $\exp (\pm 2 \wt h_{int})$. 
Then we use Jensen's inequality (\ref{one}) and use the fact that
$x\mapsto |x-y|^{- 2(1+\ep)\phi_\pm}$ is locally integrable
since $\phi_\pm < (1-\ep)$.

\bigskip

Step 6. Now we complete 
the proof of (\ref{A2}) for all squares $Q\subset K$. Since every
square can be covered by a doubled dyadic square of comparable
size, we can assume that $Q$ is such a square. If the scale of $Q$
is smaller than $M$, then $|Q|^{-1}\leq 4^{M(\mu,\ep)}$
and  we can simply use 
$e^{\pm 2h}\in L_{loc}^1$
to estimate the integrals. 

\bigskip

Step 7. Finally, we prove (\ref{errorh}). Let $\wh Q(u):= [u-1, u+1]^2$
and we split the measure $\mu_2$ as
$$
	\mu_2 = \mu_2^{int} +  \mu_2^{ext}
	:= {\bf 1}_{\wh Q(u)} \mu_2 + {\bf 1}_{\wh Q(u)^c}  \mu_2 
$$
and the function $h_2= h_2^{int} + h_2^{ext}$, where
$$
	h_2^\# (x) = {1\over 2\pi}\int_{\bR^2} 
	\log  {|x-y|\over \langle y\rangle } \mu_2^\#(\rd y) \; , 
	\qquad \# = \mbox{int, ext} \; .
$$
Similarly to the estimates (\ref{av}) and (\ref{one}) in Step 4,
we obtain
$$
	\int_{Q(u)} e^{\pm2 h_2} \leq C(\ep)
	\exp{ \Big(\pm 2\Av_{Q(u)} h_2^{ext}\Big)}
	\exp{ \Big(2 |\mu_2|(\wh Q(u)) \log \langle u \rangle\Big)} \; ,
$$
and a simple calculation shows
$$
	\Big|  \Av_{Q(u)} h_2^{ext}\Big|
	\leq \int_{Q(u)}
	\int_{\wh Q(u)^c}
	\Big| \log {|x-y|\over \langle y\rangle}\Big|
	\; |\mu_2^{ext}|(\rd y) \rd x
	\leq  |\mu_2|(\wh Q^c(u))\log \langle u \rangle + C(\ep) \; .
$$
From these estimates (\ref{errorh}) follows
using that $  |\mu_2|(\bR^2)\leq \ep$.
$\,\,\,\Box$
\bigskip

\subsection{Definition of the Pauli operator for measure valued
fields}\label{sec:core}

For any real valued function $h\in L^1_{loc}(\bR^2)$ we define the
following  quadratic form:
$$
        \cp^h(\psi, \xi): =  \cp^h_+(\psi_+, \xi_+)
	+ \cp^h_-(\psi_-, \xi_-)
$$
with
$$
	\cp^h_+(\psi_+, \xi_+):=
	4\int \overline{\partial_{\bar z} (e^{-h}\psi_+)} 
	\partial_{\bar z} (e^{-h}\xi_+) e^{2h}\; ,
	\qquad
	\cp^h_-(\psi_-, \xi_-):= 4\int
	\overline{\partial_{ z} (e^{h}\psi_-)} 
	\partial_{z} (e^{h}\xi_-) e^{-2h}
$$
on the natural maximal domains
$$
	\cD(\cp^h_\pm) = \Big\{ \psi_\pm
	\in L^2(\bR^2)\; : \; 
	\cp^h_\pm(\psi_\pm, \psi_\pm)<\infty \Big\}\; ,
$$
$$
	\cD(\cp^h) =\cD(\cp^h_+)\otimes\cD(\cp^h_-)=
	 \Big\{ \psi = \pmatrix{\psi_+\cr \psi_-}
	\in L^2(\bR^2, \bC^2)\; : \; 
	\cp^h(\psi, \psi)<\infty \Big\}\; .
$$

We use $\| \cdot \|$ to denote the usual $L^2(\bR^2, \rd x)$
or $L^2(\bR^2, \bC^2,\rd x)$ norms.
We define the following norms on functions
$$
	\triple f \triple_{h,+} := \Big[  \| f \|^2 
	+ \| \partial_{\bar z} (e^{-h}f)
	e^{h}\|^2 \Big]^{1/2}, \qquad 
	\triple f \triple_{h,-}:= \Big[  \| f \|^2 
	+ \| \partial_{z} (e^{h}f)
	e^{-h}\|^2\Big]^{1/2} 
$$
and for a spinor $\psi$  we let
\be
	 \triple \psi \triple_h: =
	\triple \psi_+ \triple_{h,+} + \triple \psi_- \triple_{h,-} \; .
\label{triplenorm}\ee

For any real function $h \in L^1_{loc}$ with $\Delta h \in \overline{\cM}$,
 we  define the set
\be
	\cC_h: = \Big\{ \psi = \pmatrix{g_+ e^{h}\cr g_- e^{-h}} 
	\; : \; g_\pm \in C_0^\infty(\bR^2)\Big\} \; .
\label{ccdef}\ee
Notice that this set
depends only on $\mu =\Delta h$: if $h, h'$ are two functions
such that $\Delta h = \Delta h' = \mu$ in distributional sense,
then $h-h'$ is harmonic, i.e., smooth. Therefore $e^h$ and $e^{h'}$
differ by a smooth multiplicative factor, i.e. $\cC_h =\cC_{h'}$,
hence we can denote this set by $\cC_\mu$.
Moreover, by Theorem \ref{thm:hdef},
for any $\mu\in \overline{\cM^*}$ and any compact set $K$,
there exists an $h \in L^1_{loc}$ with $\Delta h =\mu$ on $K$,
and  $h$ is unique modulo adding a smooth (harmonic) function.
Since the support of $g_\pm$ is compact, the following 
set is well-defined for all $\mu \in \overline{\cM^*}$
\be
	\cC_\mu : = \Big\{ \psi = \pmatrix{g_+ e^{h}\cr g_- e^{-h}} \; : \;
	 g_\pm \in C_0^\infty(\bR^2), \; \Delta h = \mu
	\quad \mbox{on} \; \mbox{supp}(g_-)\cup\mbox{supp}(g_+) \Big\} \; .
\label{cmudef}\ee

\medskip

\begin{theorem}\label{Hdef} 
Let $h\in L^1_{loc}(\bR^2)$
 be a real valued function such that $\mu:=\Delta h \in \overline{\cM^*}$.
Then 

(i) The quadratic form $\cp^h$ is  nonnegative, symmetric and closed, hence
it defines a unique self-adjoint operator $H_h$
$$
	(H_h\psi, \xi) := \cp^h(\psi, \xi)\; , \qquad \psi\in \cD(H_h), \;
	\xi\in  \cD(\cp^h)
$$
with domain
$$
	\cD(H_h) : = \{ \psi\in \cD(\cp^h)\; : \; \cp^h(\psi, \cdot)\in 
	L^2(\bR^2, \bC^2)'\}
$$

(ii) The  set $\cC_\mu$ is dense in $\cD(\cp^h)$ with respect to
 $\triple \cdot \triple_h$,
i.e., it is a form core of $H_h$.

(iii) For any  $L^1_{loc}$-functions
 $h$ and  $h'$ with $\Delta h = \Delta h' \in \overline{\cM^*}$,
the operators $H_h$ and $H_{h'}$ are unitarily equivalent
by a $U(1)$-gauge transformation. In particular, the spectral
properties of $H_h$ depend only on  $\mu=\Delta h$.
\end{theorem}

\begin{definition}\label{deff}
 For any real function $h\in L^1_{loc}$ with $\mu=\Delta h \in 
\overline{\cM^*}$
the operator $H_h$ will be called the {\bf Pauli operator
with generating potential $h$}.
For any $\mu\in \overline{\cM^*}$ 
the unitarily equivalent operators $\{ H_h\; : \;
\Delta h =\mu\}$ are called the {\bf Pauli operators 
 with a magnetic field $\mu$}. 
The Pauli operators for any $\mu\in \overline{\cM}$ are defined
as $U_\mu^*HU_\mu$ on the core $U_\mu^*C_\mu$,
 where $H$ is a Pauli operator 
with the reduced 
field $\mu^*\in\overline{\cM^*}$ (see Definition \ref{mustdef})
 and $U_\mu$ is defined in (\ref{umudef}).
\end{definition}

To complete the definition of the Pauli operator for
any magnetic field $\mu\in \overline{\cM}$, we need

\begin{theorem}\label{bm}
For any $\mu\in \overline{\cM^*}$, there exists $h\in L^1_{loc}$
with $\Delta h = \mu$ (in fact, for any $p<2$ one can find
an $h\in W^{1,p}_{loc}$). Hence the above definition of 
 $H_h$ actually defines the Pauli operators
for any measure valued magnetic field $\mu\in \overline{\cM}$.
\end{theorem}

\medskip

{\it Proof of Theorem \ref{Hdef}.} From Corollary \ref{w11}
we know that $e^{\pm h}\in 
 L^2_{loc}$, and we show below that
 for any doubled dyadic square $Q_0$ the estimate
\be
	\Big( {1\over |Q|}\int_Q e^{2h}\Big)
	\Big( {1\over |Q|}\int_Q e^{-2h}\Big) \leq C_3(h,Q_0)
\label{A2loc}\ee
analogous to (\ref{A2}) is valid  on any square $Q\subset Q_0$, 
 with a $(h, Q_0)$-dependent constant. 
These are the two properties of $h$ which we use below.

For any $Q_0$ one can find a compact set $K$ such that
$Q_0\subset \mbox{int}(K)$ and $\mu = \Delta h$
restricted to $K$, $\mu|_K$, has finite total variation.
Let $\ep =\ep(\mu|_K)/2$  and we consider
$h^{(\ep)}$ defined in Theorem \ref{thm:hdef}. Since 
$\Delta h = \Delta h^{(\ep)}$
on $K$,
we can write $h= h^{(\ep)} + \varphi$ with a smooth real function
$\varphi$ depending on $h$. In particular, for any doubled
dyadic square $Q_0$ the estimate (\ref{A2}) for $h^{(\ep)}$ implies that
(\ref{A2loc}) is valid for $h= h^{(\ep)} + \varphi$ on any square 
$Q\subset Q_0$.

\medskip

{\it Part (i).} Let $\psi_n =(\psi_{n+}, \psi_{n-})$ be a Cauchy
sequence in the norm $\triple \cdot \triple_h$,
i.e., $\psi_n\to \psi$ in $L^2(\rd x)$,
 $ \partial_{\bar z} (e^{-h}\psi_{n+}) \to u_+$
in $L^2(e^{2h}\rd x)$ and $ \partial_{ z} (e^{h}\psi_{n-}) \to u_-$
in $L^2(e^{-2h}\rd x)$.
 We have to show that
$ \partial_{\bar z} (e^{-h} \psi_+) = u_+$,
$ \partial_{z} (e^{h} \psi_-) = u_-$.
For any $\phi \in C_0^\infty (\bR^2)$ 
$$
	\int \overline{\phi} u_+ = \lim_{n\to\infty} 
	\int \overline{\phi} \partial_{\bar z} (e^{-h}\psi_{n+})
	= - \lim_{n\to\infty} 
	\int (\partial_{\bar z}\overline{\phi}) \, e^{-h}\psi_{n+}
	= -\int (\partial_{\bar z}\overline{\phi})  \, e^{-h}\psi_+
$$
hence $ \partial_{\bar z} (e^{-h}\psi_+) = u_+$ in distributional sense.
Here we used that
$$
	\Big|\int \overline{\phi} \Big( u_+
	 -  \partial_{\bar z} (e^{-h}\psi_{n+})\Big)\Big|
	\leq \|  \overline{\phi} e^{-h}\| \Big\|
	u_+ -  \partial_{\bar z} (e^{-h}\psi_{n+})\Big\|_{L^2(e^{2h})} \to 0
$$
and
$$
	\Big|\int \partial_{\bar z}\overline{\phi}\,  e^{-h}(\psi_+-\psi_{n+})
	\Big|\leq \Big\| \partial_{\bar z}\overline{\phi} e^{-h}\Big\|
	\| \psi_+-\psi_{n+}\|\to 0\; ,
$$
which follows from $e^{-h} \in L^2_{loc}$. The proof of the spin-down
component is similar. This shows that the form $\pi^h$ is closed.
The rest of the argument is standard  (see, e.g., Lemma 1 in \cite{LS}).

\bigskip

{\it Part (ii).} The spin-up and spin-down parts can be treated separately
and analogously, so  we focus only on the spin-up part. 

\medskip

Step 1. We first show that the set
$$
	\cC_0: = \{ f\in \cD(\pi^h_+), \;\mbox{supp}(f) \; \mbox{compact}\}
$$
is dense in $\cD(\pi^h_+)$ with respect to $\triple \cdot \triple_{h,+}$.
This is standard: 
let $\chi(x)$ be a compactly supported smooth cutoff function,
 $0\leq \chi \leq 1$, $\chi(x) \equiv 1$
for $|x|\leq 1$, and
let $\chi_n(x): =\chi(x/n)$. For any $f\in \cD(\pi^h_+)$ we consider
$f_n=\chi_n f$, then clearly $\triple f-f_n\triple_{h,+}\to 0$.

\bigskip

Step 2. We need the following

\begin{lemma}\label{nabla}
Let  $f\in\cC_0$ then 
 $\nabla (fe^{-h})\in L^2(e^{2h})$.
\end{lemma}

{\it Proof of Lemma \ref{nabla}.} Let $g:=fe^{-h}$.
Let $Q_1$ be a doubled dyadic square that contains a neighborhood of $K:=
\mbox{supp}\, (g)$, and let $Q_0$ be a doubled dyadic square
that strictly contains $Q_1$ and $|Q_0|=4|Q_1|$.
We define
\be
	\omega (x): = \left\{ 
	\begin{array}{cr} e^{2h(x)}& \mbox{for}\;  x\in Q_1 \cr
	1 & \mbox{for} \; x\in Q_1^c \; .
	\end{array}
	\right.
\label{omgive}\ee
\begin{lemma}\label{local}
The function $\omega (x)$ satisfies the  inequality
\be
	\Big( {1\over |Q|}\int_Q \omega\Big)
	\Big( {1\over |Q|}\int_Q \omega^{-1}\Big) \leq C_4(h, Q_0)
\label{hold}\ee
for any square $Q\subset \bR^2$, i.e.,
$\omega$ is an $A_2$-weight (see \cite{GR, St}).
\end{lemma}

{\it Proof of Lemma \ref{local}.} It is sufficient to prove (\ref{hold}) for
all doubled dyadic squares $Q$. 
It is easy to see that one of the following cases occurs:
(i) $Q$ is disjoint from $Q_1$,
(ii) $Q\subset Q_0$, (iii) $|Q_1| \leq 9 |Q|$.
In the first case
 (\ref{hold}) is trivial, in case (ii)
it follows from (\ref{A2loc}).
Finally, in case (iii) we have
$$
	\Big( {1\over |Q|}\int_Q \omega\Big)
	\Big( {1\over |Q|}\int_Q \omega^{-1}\Big)
	\leq 36^2
        \Big( 1 + {1\over |Q_0|}\int_{Q_0} e^{2h}\Big)
	\Big( 1 + {1\over |Q_0|}\int_{Q_0} e^{-2h}\Big)
$$
hence (\ref{hold}) holds with an appropriate constant.
 $\,\,\Box$.

\medskip

Since
$|\nabla g|^2 = 2(|\partial_z g|^2 + |\partial_{\bar z} g|^2)$
and $\omega = e^{2h}$ on $\mbox{supp}\, (g)$,
Lemma \ref{nabla} follows immediately from 
\be
	\int_{\bR^2} |\partial_{z}g|^2\omega
	\leq C_5(h, Q_0)
	 \int_{\bR^2} |\partial_{\bar z}g|^2\omega \, .
\label{key1}
\ee
Notice that
$$
	\wh{\partial_{ z}g} (\xi) = m(\xi)\wh{\partial_{ \bar z}g} (\xi)
	\quad \mbox{with} \quad m(\xi):=
	{(\xi_1 -i\xi_2)^2\over |\xi|^2}\; ,
$$
where hat stands for Fourier transform, $\xi\in \bR^2$,
and $m(\xi)$ is a homogeneous
multiplier of degree 0. Hence
(\ref{key1}) is just the weighted $L^2$-inequality
for the regular singular integral operator $T_m$ with Fourier multiplier 
$m(\xi)$ and with weight $\omega\in A_2$ \cite{GR, St}. $\,\,\Box$

\bigskip

Step 3. To conclude that $\cC_0\cap e^h C_0^\infty$ is dense
in $\cC_0$ with respect to $\triple \, \cdot\, \triple_{h,+}$,
 we  use the fact that $C_0^\infty$ is dense in the
weighted Sobolev space $W^{1, 2}(\omega)$
with the $A_2$-weight $\omega$ (see e.g. \cite{K}).
Here we only recall the key point of the proof.
Let $g\in W^{1, 2}(\omega)$ compactly supported and 
 $g_\ep := J_\ep \ast g \in C_0^\infty$ where
$J_\ep(x): = \ep^{-2} J(x/\ep)$ is a standard mollifier: $0\leq J \leq 1$,
$\int J =1$, $J$ smooth, compactly supported. 
Then the functions $|\nabla g_\ep| \leq   J_\ep \ast |\nabla g|$, 
 have an $L^2$-integrable majorant by the weighted
maximal inequality \cite{St} applied to $|\nabla g| \in L^2(\omega)$,
hence $g_\ep\to g$ in $W^{1, 2}(\omega)$ as $\ep\to0$.
Notice that every $g_\ep$ is supported on a common
compact neighborhood of the support of $g$.

\bigskip

{\it Part (iii).}
Since $\Delta h = \Delta h'$, we can write $h' = h +\varphi$
with a smooth real function $\varphi$.
We  define $\lambda$ as the harmonic conjugate of $\varphi$,
 $\nabla \lambda = \nabla^\perp\varphi$,
which exists and is smooth by $\Delta \varphi =0$. By
 $\partial_{\bar z} (\varphi + i\lambda)=0$ we have
$$
	\pi^h(\psi, \psi) = \pi^{h'}\Big( e^{-i\lambda}\psi,
	 e^{-i\lambda}\psi\Big) \; ,\qquad \psi \in \cC_\mu \; ,
$$
 and then by the density of $\cC_\mu$
we obtain the same relation for all $\psi\in \cD(\pi^h)$.
$\,\,\,\Box$

{\it Proof of Theorem \ref{bm}.} 
Since $|\mu|$ is finite on every bounded set, we can find
a sequence of disjoint
rings, $R_j := \{ x\; : \; r_j \leq |x| \leq r_j+2\delta_j\}$,
 $j=1, 2,\ldots$,
with appropriate widths $2\delta_j>0$ and
radii $r_j\to\infty$ (as $j\to\infty$), such
that $\sum_j |\mu|(R_j) < \infty$. For  $j=0$ we set $r_j=\delta_j=0$.
 Let $0\leq \chi_j\leq 1$ ($j=0,1,\ldots$)
 be smooth functions such that
$\chi_j(x)\equiv 1$ for $r_j +2\delta_j \leq |x|\leq r_{j+1}$ and
$\chi_j(x) \equiv 0$ for $|x|\leq r_j +\delta_j$
 or $|x|\ge r_{j+1}+\delta_{j+1}$.
Notice that the supports of $\chi_j$ are disjoint.

We define $\mu_j:=\mu \cdot {\bf 1}\{ x\; : 
r_j+2\delta_j \leq |x|\leq r_{j+1}\}$.
 By Theorem \ref{thm:hdef} 
there exist $h_j \in \bigcap_{p<2}W^{1,p}_{loc}$, 
$e^{\pm h_j}\in L^2_{loc}$
 with  $\Delta h_j = \mu_j$. 
We notice that
$$
	\Delta \Big(\sum_j \chi_j h_j) =
	\nu+\sum_j \chi_j \mu_j
$$
where $\nu$ is absolutely
continuous, $\nu = N(x)\rd x$ with 
$$
	N =  \sum_j \Big[ 2\nabla \chi_j\cdot \nabla h_j
	+ h_j \Delta \chi_j\Big] \in L^1_{loc}
$$
We can find a decomposition $N= N_1 + N_2$, $N_1\in L^\infty_{loc}$ and
$N_2\in L^1(\bR^2)$.

Let $\kappa : = \mu - \sum_j \chi_j \mu_j - N_2(x) \rd x$,
then $\kappa\in \cM$ since $N_2\in L^1(\bR^2)$ and
the measure $\mu - \sum_j \chi_j \mu_j$ belongs to $\cM$ since it vanishes
on the complement of $\bigcup_j R_j$, it
has a total variation  smaller than $|\mu|$ on  each $R_j$
and $\sum_j |\mu|(R_j)<\infty$. It is also clear that
$\kappa$ does not charge more to any point than $\mu$ does
since $0\leq\chi_j\leq 1$ and they have disjoint supports,
hence $\kappa\in \cM^*$.

By Theorem \ref{thm:hdef} there is
 $k  \in \bigcap_{p<2}W^{1,p}_{loc}$, 
$e^{\pm k}\in L^2_{loc}$ such that $\Delta k =\kappa$.
We define $h^*: = k + \sum_j \chi_j h_j$, clearly 
$h^* \in \bigcap_{p<2}W^{1,p}_{loc}$ and
$\mu = \Delta  h^* -N_1(x)\rd x$.

By the Poincar\'e formula
there exists $\bA\in L^\infty_{loc}$
with $\nabla^\perp\!\cdot \! \bA = - N_1$.
For any fixed $p<2$ (even for $p<\infty$), 
 one can find $\wt\bA\in L^p_{loc}$
with $\nabla^\perp\!\cdot \! \wt\bA  = - N_1$, $\nabla\cdot \wt\bA =0$
by Lemma 1.1. (ii) \cite{L}. But then $\wt\bA^\perp = (-\wt A_2, \wt A_1)$
is curl-free, hence $\wt\bA^\perp = \nabla \wt h$ for some $\wt h \in
W^{1,p}_{loc}$ by Lemma 1.1. (i) \cite{L}. Then $\Delta \wt h = N_1$,
hence  $h:= h^* - \wt h$
 satisfies $\Delta h =\mu$ and  we see that 
$h\in W^{1,p}_{loc}\subset L^1_{loc}$.
$\;\;\Box$

\medskip

Finally, we have to verify that the Pauli operator $H_h$
defined in this Section coincides with the standard Pauli
operator if $\bA\in L^2_{loc}$, modulo a gauge transformation.

\begin{proposition}\label{coinc}
Let $\bA\in L^2_{loc}$.
We assume that $\nabla^\perp\!\cdot\!  \bA$ (in distributional sense)
is a measure and that  $\mu:= \nabla^\perp\!\cdot\! \bA\in \overline{\cM}$.
Then $\mu\in \overline{\cM^*}$, in fact $\mu$ has no discrete component.
Moreover,  if $\Delta h =\mu$ with some $h\in L^1_{loc}$, then
the operator $H_h$ defined in Theorem \ref{Hdef}
is unitarily equivalent to the Pauli operator $P_\bA^{max}$ 
associated with the maximal form $p_\bA$ on $\cD_{max}(p_\bA)$ as 
defined in Section \ref{stand}. 
\end{proposition}

{\it Remark:}  $\bA\in L^2_{loc}$ does not imply that
$\nabla^\perp\!\cdot\! \bA$ is even locally a measure of finite variation.
One example is the radial gauge $\bA (x): = \Phi(|x|)|x|^{-2} x^\perp$,
$x^\perp:=(-x_2, x_1)$,  that generates the
radial field 
$$
	B(x): = \sum_{n=1}^\infty {(-4)^{n}\over n}\cdot
	{\bf 1}( 2^{-n} \leq |x| < 2^{-n+1}).
$$
with flux $\Phi(r): = \int_{|x|\leq r} B(x) \rd x$.
 One can easily check that
$\int_{|x|\leq 1} |B(x)| \rd x = \infty$ but
$\bA\in L^2_{loc}$.
However, if $\nabla^\perp\!\cdot\!   \bA\ge 0$ as a distribution, then
it is a (positive) Borel measure $\mu \in \overline{\cM^*}$
 (see \cite{LL}).

\bigskip

{\it Proof.} First we show that $\mu = \nabla^\perp\!\cdot\!  \bA$
has no discrete component. 
Suppose, on the contrary, that $\mu(\{ x\})\neq 0$  for some $x$,
and we can assume $x=0$, $\mu(\{ 0 \})>0$. 
Let $\chi$ be a radially symmetric smooth function on $\bR^2$,
$0\leq \chi \leq 1$, $\mbox{supp}\chi \subset \{ |x|\leq 2\}$,
$\chi (x)\equiv 1$  for $|x|\leq 1$,
$|\nabla \chi|\leq 2$, and let $\chi_n(x) : = \chi(2^n x)$.
Clearly $-\int \bA\cdot\nabla^\perp \chi_n  =\int \chi_n \rd \mu \to 
\mu(\{ 0\})$ as $n\to\infty$. Using polar coordinates,
we have,  for large enough $n$,
$$
	{1\over 2}\mu(\{ 0 \}) \leq - \int \bA\cdot \nabla^\perp \chi_n
	\leq 4\int_{2^{-n}}^{2^{-n+1}} \int_0^{2\pi}|\bA(s, \theta)| 
	\rd \theta\, \rd s
$$
$$
	\leq 4\sqrt{2\pi}
	 \Bigg(\int_{2^{-n}}^{2^{-n+1}} \int_0^{2\pi}|\bA(s, \theta)|^2 
	\rd \theta\, s \, \rd s\Bigg)^{1/2} 
	\leq  4\sqrt{2\pi}
	 \Bigg(\int |\bA(x)|^2  \cdot {\bf 1}(2^{-n}\leq |x|\leq 2^{-n+1})
	\rd x\Bigg)^{1/2} 
$$
hence $\int_{|x|\leq 1} |\bA|^2=\infty$. 
The proof also works if we assume only $\mu \in \overline{\cM}$
instead of $\mu \in {\cM}$.

\medskip
Now we prove the unitary equivalence. Without loss of generality
we can assume that $\nabla\cdot \bA =0$
by part (ii) Lemma 1.1. of \cite{L}.
 Let $\bA_h: = \nabla^\perp h$, 
then $\nabla^\perp\!\cdot\!   \bA_h = \mu$,
 $\nabla\cdot \bA_h =0$ and $\bA_h\in L^1_{loc}$  by Corollary \ref{w11}. 
Since  $\nabla^\perp\!\cdot\!  (\bA- \bA_h) = 0$,
 there exists $\lambda \in W^{1,1}_{loc}$ such that $\bA = \bA_h + \nabla
\lambda$ by part (i)  Lemma 1.1. of \cite{L}. Taking the divergence,
we see that $\Delta \lambda =0$, hence $\lambda$ is smooth.
 Let $\varphi$ be a smooth
harmonic conjugate of $\lambda$, $\nabla\lambda = \nabla^\perp \varphi$.
We have the following identity
$$
	p_\bA(\psi, \psi)
	= \pi^{h+\varphi}(\psi, \psi) 
	=\pi^h \Big( e^{i\lambda}\psi, e^{i\lambda}\psi\Big)\; .
$$
From the first equality we obtain
that  $\cD_{max}(p_\bA) = \cD(\pi^{h+\varphi})$, i.e., $P_\bA^{max}=
H_{h+\varphi}$ on $\cD(P_\bA^{max})=
\cD(H_{h+\varphi})$. From the second equality
it follows that $H_{h+\varphi} = e^{-i\lambda} H_h e^{i\lambda}$
and that $\cD(H_{h+\varphi}) = e^{-i\lambda} \cD(H_h)$.
In fact, $\cD(H_h) = \cD(H_{h+\varphi})$ since
the multiplication by the smooth factor $e^{i\lambda}$ leaves the form core
$\cC_h = \cC_{h+\varphi} = \cC_\mu$ invariant.
 $\;\;\;\Box$

\subsection{Pauli operator generated by both potentials}\label{sec:inf}

 Theorem \ref{bm}  showed that every measure $\mu\in \overline{\cM^*}$
can be generated by an $h$-potential, $\Delta h =\mu$,
and we defined the Pauli operators.
However, it may be useful to combine the scalar potential
with the usual vector potential $\bA\in L^2_{loc}$ to generate
the given magnetic field. In this way one has more freedom in
choosing the potentials.
Typically, the singularities can be easier handled by
the $h$-potential, and the standard $h=\sfrac{1}{2\pi}
\log |\, \cdot \, |\ast\mu$
formula is (locally) available. But this formula exhibits a strong 
non-locality of $h$, and the truncation method of the proof
of Theorem \ref{bm} is not particularly convenient in practice.
Large distance behavior of the bulk magnetic field is better
described by a vector potential. 
In this section we give such a unified definition of the Pauli
operator.

\bigskip

For any  $h\in L^1_{loc}$, $\bA\in L^2_{loc}$ we define
the quadratic form
$$
	\pi^{h,\bA}(\psi, \psi)
	: = \int \Big| (-2i\partial_{\bar z} + a)(e^{-h}\psi_+)\Big|^2
	e^{2h}
	+ \int \Big| (-2i\partial_{z} + \bar{a})(e^{h}\psi_+)\Big|^2e^{-2h}
$$
on the maximal domain
$$
	\cD(\pi^{h, \bA}) : = \Big\{ \psi \in L^2(\bR^2, \bC^2)
	\; : \; \triple \psi \triple_{h,\bA} <\infty\Big\}\; ,
$$
where $a= A_1+ iA_2$ and
$$
	\triple \psi \triple_{h,\bA}: =
	\Big[ \| \psi \|^2 + \pi^{h,\bA}(\psi, \psi)\Big]^{1/2} \; .
$$

Let
$$
	\cP^*: = \Big\{ (h, \bA) \; : \; h\in L^1_{loc}, 
	\; e^{\pm h}\bA \in L^2_{loc}, \;
	\Delta h \in \overline{\cM^*},
	 \nabla^\perp\!\cdot\!  \bA \in \overline{\cM}\Big\}
$$
be the set of admissible potential pairs. The measure
$\mu : = \Delta h + \nabla^\perp\!\cdot\!  \bA \in \overline{\cM}$ is
called the magnetic field generated by $ (h, \bA)$. 
We recall from Corollary \ref{w11} 
that $(h, \bA)\in\cP^*$ implies $h\in \bigcap_{p<2} W^{1,p}_{loc}$
 and $e^{\pm h} \in L^{2}_{loc}$,
moreover, $e^{\pm h} \bA\in L^2_{loc}$ implies $\bA\in L^2_{loc}$.

Since $\nabla^\perp\!\cdot\!  \bA $ has no discrete component
(Proposition \ref{coinc}), the measure $\mu$ generated by
 $ (h, \bA)\in \cP^*$ is in $\overline{\cM^*}$.
In particular, the set of measures generated by 
a potential pair from $\cP^*$ is the same as the 
set of measures generated by only $L^1_{loc}$ $h$-potentials
(Theorem \ref{bm}).

\medskip

\begin{theorem}\label{thm:inf}
(i) (Self-adjointness). Assume that  $ (h, \bA)\in \cP^*$
and let $\mu: =\Delta h + \nabla^\perp\cdot \bA$.
Then  $\pi^{h, \bA}$ is a nonnegative symmetric
 closed form, hence it defines a
unique self-adjoint operator $H_{h,\bA}$. 

(ii)  (Core). The set 
$\cC_{\mu}$ (see (\ref{cmudef}))  is dense in $\cD(\pi^{h,\bA})$
with respect to $ \triple \, \cdot \, \triple_{h,\bA}$, i.e.,
it is a form core for  $H_{h,\bA}$.

(iii) (Consistency). If  $ (h, \bA)\in \cP^*$ and $\wt h\in L^1_{loc}$
such that $\Delta h + \nabla^\perp\!\cdot\!\bA = \Delta \wt h$, then
$H_{h,\bA}$ is unitary equivalent to $H_{\wt h}$ defined
in Theorem \ref{Hdef}.

\end{theorem}

\begin{definition} For any $(h,\bA)\in \cP^*$ the operator
$H_{h,\bA}$ is called the {\bf Pauli operator with a potential 
pair $(h, \bA)$}.
\end{definition}

Notice that Proposition \ref{coinc} and (iii) of  Theorem \ref{thm:inf}
guarantees that the Pauli operators with the same magnetic
field are unitarily equivalent, irrespectively which definition
we use.

{\it Proof of Theorem \ref{thm:inf}.} {\it Part (i).}
The proof that $\pi^{h, \bA}$ is closed   is very similar
 to  the proof
of part (i)  of Theorem \ref{Hdef}.
 The operators $\partial_{\bar z}$
and $\partial_{z}$ should be replaced by  $\partial_{\bar z}+ia$
and $\partial_{z}-i\overline{a}$, but the extra terms with 
$a$ can always be estimated by the local $L^2$ norm of $e^{\pm h}\bA$.

{\it Part (ii).} Step 1.
We need the following preliminary observation. Since
$\bA\in L^2_{loc}$,
we can consider the decomposition $\bA= \wt\bA + \nabla\wt\lambda$,
$\nabla\cdot\wt\bA=0$, $\wt\bA\in L^2_{loc}$, $\wt\lambda\in W^{1,2}_{loc}$
(see Lemma 1.1 \cite{L}).
 $\wt\bA$ and $\wt\lambda$ are called the divergence-free
and the gradient component of $\bA\in L^2_{loc}$, and notice
that  $\wt \bA$ is  unique up to a smooth gradient,  since
if $\wt \bA + \nabla\wt\lambda =\wt \bA' + \nabla\wt\lambda'$
then $0= \nabla\cdot(\bA-\bA')= \Delta(\wt\lambda' - \wt\lambda)$,
i.e., $\wt\lambda' - \wt\lambda$ is smooth.  

Moreover, if $\bA e^{\pm h}\in L^2_{loc}$, then $\wt\bA 
e^{\pm h}\in L^2_{loc}$ as well. To see this,
we fix a compact set $K$ and  a compact set $K^*$
whose interior contains $K$, then we
 choose a cutoff function $0\leq\varphi_K\leq 1$
with $\varphi_K\equiv 1$ on $K$ and $\mbox{supp}\, \varphi_K \subset K^*$.
We let $\bA_K: = \varphi_K\bA \in L^2$
and let $\lambda$ be defined via its Fourier transform
$$
	\wh \lambda (\xi): = {\xi \cdot \wh\bA_K(\xi)\over |\xi|^2}\; ,
	\qquad\xi\in \bR^2
$$
i.e., $-\Delta \lambda = \nabla\cdot \bA$.
Then $\nabla \lambda$ is obtained from $\bA_K$ by
the action of a  singular integral operator
whose multiplier is $\xi\otimes \xi/|\xi|^2$.
Choose $\omega$ as in (\ref{omgive}),
 where $Q_0$ is a dyadic square containing $K^*$,
then $\omega\in A_2$.
Hence, by the weighted $L^2$-inequality we have
$$
	\int |\nabla \lambda|^2 \omega \leq C(\omega) \int |\bA_K|^2 \omega
	= C(\omega)\int |\bA_K|^2  e^{2h}
$$
with some
$\omega$-dependent constant. In particular 
$\nabla\lambda \in L^2_{loc}(e^{ 2h})$. The proof
of $\nabla\lambda \in L^2_{loc}(e^{ -2h})$ is identical.
Now $\wt\lambda$ satisfies $\Delta \wt \lambda = -\Delta \lambda$
on $K$, i.e. $\wt\lambda$ and $\lambda$ differ by an additive
smooth function, hence $\nabla\wt\lambda \in L^2_{loc}(e^{\pm 2h})$,
which means that $\wt\bA 
e^{\pm h}\in L^2_{loc}$.

\medskip

Step 2. We show  that $\cC_\mu \subset\cD(\pi^{h, \bA})$
if $(h, \bA)\in \cP^*$, $\mu=\Delta h +\nabla^\perp\cdot \bA$.
Let $\psi \in \cC_\mu$ be compactly supported on $K$ and let $K^*$ be
a compact set whose interior contains $K$.
 Since
$\mu$ restricted to $K^*$ has finite total variation, we can apply
 Theorem \ref{thm:hdef} for the restricted measure to construct a
function $h^*\in L^1_{loc}$ such that $\Delta h^* = \mu = \Delta h
+ \nabla^\perp\cdot\wt\bA$ on $K^*$. Then
there is a real function $\chi$ such that $\wt\bA = \nabla^\perp (h^*-h) +
\nabla\chi$ (Lemma 1.1. of \cite{L}).
After taking the divergence, we see that
$\chi$ is harmonic on $K^*$.
 Let $\varphi\in C^\infty$ be its harmonic conjugate,
$\nabla\chi=\nabla^\perp\varphi$. We have the identity
\be
	\pi^{h, \bA}(e^{-i\wt\lambda}\psi, e^{-i\wt\lambda} \psi) = 
	\pi^{h, \wt\bA}(\psi, \psi)
	=\pi^{h^*+\varphi}\Big( \psi,  \psi\Big)
\label{gaid1}\ee
for any $\psi$ supported on $K^*$. Since $\psi  \in \cC_\mu$,
we can write $\psi_\pm = g_\pm e^{\pm h^*}$ and we see that
the right hand side of (\ref{gaid1}) 
is finite, hence $e^{-i\wt\lambda}\psi \in
\cD(\pi^{h, \bA})$. But by Schwarz inequality
$$
	\pi^{h, \bA}(e^{-i\wt\lambda}\psi, e^{-i\wt\lambda} \psi)
	 \ge {1\over 2} \pi^{h, \bA}(\psi, \psi)
	- 4 \sum_\pm
	\|g_\pm\|_\infty^2\int_{K^*} |\nabla\wt\lambda|^2 e^{\pm 2h}
$$
hence $\psi \in \cD(\pi^{h, \bA})$ by Step 1.

\medskip

Step 3. We now show that $\cC_\mu$ is dense in $\cD(\pi^{h, \bA})$
with respect to $\triple \, \cdot \,\triple_{h, \bA}$
if $(h, \bA)\in \cP^*$, $\mu=\Delta h +\nabla^\perp\cdot \bA$.
We first notice that it is sufficient to show that 
$\cC_{\mu}$ is dense in the set 
$$
	\overline{\cC}_0:=\{ \psi \in \cD(\pi^{h, \bA}), \;
	: \;\mbox{supp}(\psi) \;\mbox{compact}\}
$$ 
similarly to Step 1
of the proof of Theorem \ref{Hdef} (ii). So let $\psi\in \cD(\pi^{h, \bA})$
be supported on a compact set $K$. As in Step 2, we let $h^*\in L^1_{loc}$ 
be a function such that $\Delta h^* = \mu$ on a compact neighborhood $K^*$
of $K$, $K\subset \mbox{int}(K^*)$. As before, we have 
 $\wt\bA = \nabla^\perp (h^*-h) + \nabla\chi$ with a harmonic $\chi$
and let  $\varphi\in C^\infty(K^*)$ be its harmonic conjugate,
$\nabla\chi=\nabla^\perp\varphi$. The identity (\ref{gaid1})
is now written as
\be
	\pi^{h, \bA}(\psi, \psi)  
	=\pi^{h^*+\varphi}\Big( e^{i\wt\lambda}\psi,  e^{i\wt\lambda}\psi\Big)
\label{gaid}\ee
for any $\psi$ supported on $K^*$. In particular, 
$\psi\in \cD(\pi^{h, \bA})$ implies $e^{i\wt\lambda}\psi\in\cD(\pi^{h^*
+\varphi})$.

We define the set
$$
	\wt\cC_\mu: = \Bigg\{ \psi =\pmatrix{g_+ e^h\cr g_-e^{-h}}
	\; : \; g_\pm \in L^{\infty}_0, \; \Delta h = \mu
	\quad \mbox{on}\; \mbox{supp} (g_-) \cup
	\mbox{supp} (g_+) \Bigg\}
$$
where $L_0^\infty$ denotes the set of bounded, compactly supported
functions. The  set $\wt\cC_\mu$ is well defined, see the remark before
the definition (\ref{cmudef}).

Since  $\cC_\mu$ is dense in  $\cD(\pi^{h^*+\varphi})$
with respect to  $\triple\, \cdot \, \triple_{h^*+\varphi}$
by  part (ii)  of Theorem \ref{Hdef}, we 
can find a sequence of spinors $\xi_n \in 
\cC_{\mu}$ such that $\triple \xi_n -  e^{i\wt\lambda}\psi
\triple_{h^*+\varphi}\to0$. We can assume that all $\xi_n$ are supported
in $K^*$ (see remark at the end of Step 3 of the proof Theorem 
\ref{thm:hdef} (ii)).  But then 
$\triple e^{-i\wt\lambda} \xi_n -  \psi
\triple_{h,\bA}\to0$ again by (\ref{gaid}), in particular the
set $\overline{\cC}_1 :=  \cD(\pi^{h, \bA})\bigcap \wt\cC_\mu$
is dense in  $\Big( \cD(\pi^{h, \bA}), 
\triple\, \cdot \, \triple_{h, \bA}\Big)$. 

Finally, we show that $\cC_\mu$ is dense in $\Big( \overline{\cC}_1,
\triple\, \cdot \, \triple_{h, \bA}\Big)$. 
 Let $\chi \in  \overline{\cC}_1$, i.e., $\chi_\pm = g_\pm e^{\pm h}$
with some compactly supported bounded functions $g_\pm$.

Notice that if $g$ is a bounded function, then $ge^{\pm h}\bA \in L^2_{loc}$
since $e^{\pm h}\bA\in L^2_{loc}$. In particular
$(\partial_{\bar z} + ia)g_+   \in L^2(e^{2h})$ 
implies  $\partial_{\bar z}g_+  \in  L^2(e^{2h})$.
But then $\nabla g_+  \in  L^2(e^{2h})$ by Lemma \ref{nabla}
and similarly for $g_-$.
We focus only on the spin-up part, 
the spin-down part is similar.

Let $g^{(\ep)} : = J_\ep\ast  g_+$,
where $J_\ep$ is a standard mollifier (see Step 3. of the proof 
of Theorem \ref{Hdef} (ii)). Recall that $\|g^{(\ep)}  \|_\infty
\leq \|g_+\|_\infty$ and the functions
$|\nabla g^{(\ep)}|\leq J_\ep\ast |\nabla g_+|$
have an $L^2(e^{2h})$-integrable majorant using the weighted
maximal inequality.
By passing to a subsequence
$g^{(\ep)}\to g_+$, $\nabla g^{(\ep)} \to\nabla g_+$
in $L^2(e^{2h})$ and $g^{(\ep)}\to g_+$  a.e. as $\ep\to0$.
Therefore
$$
	 \int \Big|(\partial_{\bar z}+ia) (g^{(\ep)}  - g_+)\Big|^2
	e^{2h} 
	 + \Big\| (g^{(\ep)}  - g_+)e^{h}\Big\|^2 
$$
$$
	\leq  2\int \Big|\nabla (g^{(\ep)}  - g_+)\Big|^2
	e^{2h} + 2\int |\bA|^2 |g^{(\ep)}  - g_+|^2
	e^{2h} 
	 + \Big\| (g^{(\ep)}  - g_+)e^{h}\Big\|^2  \to 0
$$
as $\ep\to0$ since
 $e^{ h}\in L^2_{loc}$ and $e^{h}\bA\in L^2_{loc}$.

\medskip

{\it Part (iii).} Since $\Delta \wt h \in \overline{\cM^*}$,
we know that $\wt h\in L^2_{loc}$ (Corollary \ref{w11}).
 Since $\nabla^\perp\cdot (\nabla^\perp h -
\nabla^\perp\wt h + \bA)=0$, there exists $\lambda\in W^{1,2}_{loc}$
with $\nabla^\perp(h-\wt h) + \bA = \nabla\lambda$.
A simple calculation shows that
$$
	\pi^{h, \bA}(\psi, \psi) = \pi^{\wt h}( e^{i\lambda}\psi,
	 e^{i\lambda}\psi) \; . \qquad \qquad \Box
$$

\section{Aharonov-Casher theorem}

We prove the following extension of the  Aharonov-Casher
theorem:

\begin{theorem}\label{ACprec} Let $\mu\in \overline{\cM}$ and 
we assume that  $\mu^*\in \cM^*$
i.e., we assume that after reducing the point masses in $\mu$
 the reduced mesure $\mu^*$ has finite total variation 
(see Definition \ref{mustdef}).
Let $\Phi: = \sfrac{1}{2\pi} \int \mu^*$.
The dimension of the kernel of any Pauli operator $H$
with magnetic field $\mu$ is given
\be
	\mbox{dim} \mbox{Ker} \, H = \left\{
	\begin{array}{ccl}
	[ |\Phi| ]  & \mbox{if} & \Phi\not\in\bZ 
	\;\; \mbox{or} \;\; \Phi=0\cr
	[ |\Phi| ] \; \mbox{or} \;  [ |\Phi| ]-1 & \mbox{if} & 
	\Phi\in\bZ \setminus\{ 0\}
	\end{array}
	\right. 
\label{acfor}
\ee
(here $[a]$ denotes the integer part of $a$).
In case of nonzero integer $\Phi$ (second line) both cases can
occur, but if, additionally, $\mu^*$ has a compact support or
has a definite sign, then always
 $\mbox{dim} \mbox{Ker} \, H = [ |\Phi| ]-1$.
In all cases the kernel is in the eigenspace of $\sigma_3$:
$\mbox{Ker}(H)\subset\{ \psi\; : \; \sigma_3 \psi = -s\psi\}$ 
with $s=\mbox{sign}(\Phi)$.
\end{theorem}

Combining this Theorem with Proposition \ref{coinc} we obtain

\begin{corollary} If $\bA\in L^2_{loc}$ and $B\in L^1$, then
the dimension of the kernel of $P_\bA^{max}$
is given by (\ref{acfor}) where $\Phi = \sfrac{1}{2\pi} \int B$.
$\;\;\;\Box$
\end{corollary}

\bigskip

{\it Proof of Theorem \ref{ACprec}.} 
We can assume that $\mu\in \cM^*$. Recalling the definition of
$\ep(\mu)$ from (\ref{epdef}) we
 apply Theorem \ref{thm:hdef} to
choose $h:= h^{(\ep)}$ with some $\ep < \ep(\mu)$.
By (iii) of Theorem \ref{Hdef} it is sufficient to
consider the operator $H_h$. We can also assume that
$\Phi\ge 0$.

Suppose first that $\Phi$ is not integer, let $\{ \Phi \} = \Phi - [\Phi]$
be its fractional part.
Choose $\ep <\min\Big\{ \ep (\mu), \{ \Phi \}/3, (1-\{ \Phi \})/3 \Big\}$. 
Any  normalized eigenspinor $\psi$ with $\pi^h (\psi, \psi)=0$
 must be in the form $\psi = ( e^{h} g_+, e^{-h} g_-)$ where
$g_+$ is holomorphic and $g_-$ is antiholomorphic.

First we show that $g_+ =0$. Let $u\in\bR^2$ with $|u|\ge R(\ep, \mu)+1$.
We use the decomposition $h=h_1+h_2$ from Theorem \ref{thm:hdef}. We have
$$
	1 = \|\psi\|^2 \ge 
	\int_{Q(u)} e^{2h}|g_+|^2 \ge
	\Big( \int_{Q(u)} e^{h_1}|g_+| \Big)^2
	\Big( \int_{Q(u)} e^{-2h_2} \Big)^{-1},
$$
and by (iii) of Theorem \ref{thm:hdef} and subharmonicity of $|g_+|$
 we see that $|g_+(u)|\leq C\int_{Q(u)}|g_+| \leq
C\langle u \rangle^{-\Phi+2\ep} \to 0$ as $u\to\infty$, hence $g_+=0$.

A similar calculation shows that 
$|g_-(u)| \leq C\langle u \rangle^{\Phi+2\ep}$, i.e., $g_-$
must be a polynomial of degree  at most $[\Phi]$ since $\Phi + 2\ep < 
[\Phi]+1$. However, a polynomial of degree $[\Phi]$ would give
$$
	\int e^{-2h} |g_-|^2 \ge C\sum_{k\in \Lambda_0, |k|\ge R}
	|k|^{2[\Phi]}	
	\int_{Q(k)} e^{-2h} \ge C\sum_{k\in \Lambda_0, |k|\ge R}
	|k|^{2[\Phi]} \Big(\int_{Q(k)} e^{2h}\Big)^{-1}
$$
$$ 
	\ge  C\sum_{k\in \Lambda_0, |k|\ge R}
	|k|^{2([\Phi]-\Phi -2\ep)} =\infty
$$
for some large enough $R$ and various constants $C$.

On the other hand, the functions $g_-(z) = 1, z, \ldots , z^{[\Phi]-1}$ all
give normalizable spinors since for these choices
\be
	\int e^{-2h}|g_-|^2 \leq C\sum_{k\in \Lambda_0}
	|k|^{2([\Phi]-\Phi-1)}	\int_{Q(k)} e^{-2h_2} 
	\leq  C\sum_{k\in \Lambda_0}
	|k|^{2([\Phi]-\Phi-1+\ep)} <\infty \; .
\label{thres}\ee

If $\Phi$ is integer, then the same arguments work except
(\ref{thres}); in fact $g_-(z)=z^{\Phi-1}$ may or may not give
normalizable solutions.

If $\mu$ is compactly supported and $\Phi\ge 0$ 
is integer then from the definition
of $h$ (\ref{hddef})-(\ref{htotdef}) we see that
$\Big|h(x) - \Phi \log|x|\Big|$ is bounded for all large enough $|x|$.
Similarly one can easily verify that if $\mu\ge 0$, then
$h(x) \leq \Phi \log \langle x \rangle + C$ since
$\log |x-y|\leq \log  \langle x \rangle  +\log  \langle y \rangle +C$.
In both cases $z^{\Phi-1}e^{-h}$ is not $L^2$-normalizable at infinity.

However, if $\mu$ can change sign and not compactly supported
then there could be $\Phi\in \bZ$ zero energy states.
For example the radial field (with $\beta >0$, $N\in {\bf N}$) 
$$
	B(x)= \left\{ 
	\begin{array}{ccc}
	2(N+\beta) e^{-2} & \mbox{for} & |x|\leq e \cr
	-\beta (|x|\log |x|)^{-2} & \mbox{for} & e <|x|
	\end{array}
	\right. \; ,
$$
with $\Phi = \sfrac{1}{2\pi} \int B =N$,
is generated by a radial potential $h(x)$
such that  $h(x) = N\log |x| + \beta \log\log |x|$
for large $x$ and is regular for small $x$. 
The threshold state
$z^{\Phi-1}e^{-h}$ is normalizable only for $\beta >\sfrac{1}{2}$,
so in this case the dimension of the kernel  is $\Phi$, otherwise $\Phi -1$.
$\,\,\,\,\Box$

\bigskip

This theorem requires $\mu\in\cM$. We will see in Section
\ref{count} that the Aharonov-Casher theorem need not be
true for magnetic fields with infinite total variation.
However, the proof above still works for magnetic fields
that can be decomposed into the sum of a component in $\cM$ and
a component with  a regularly behaving generating potential.
We just remark one possible extension:

\begin{corollary}\label{reg} Suppose that $\mu\in \overline{\cM}$ 
can be written as $\mu = \mu_{rad} + \wt\mu$ such that
 $\wt \mu\in \cM$ and $\mu_{rad}$ is a rotationally symmetric Borel
measure (i.e., $\mu_{rad} = \mu_{rad}\circ R$ for any rotation $R$ in $\bR^2$
around the origin). We can assume that $\mu_{rad}(\{ 0 \}) =0$
by including the possible delta function at the origin into $\wt\mu$.
 We assume that
that $\Phi_{rad}: =\lim_{R\to\infty} \Phi(R):=
\lim_{R\to\infty}\sfrac{1}{2\pi} \int_{|x|\leq R} \mu_{rad}(\rd x)$ exists 
(possibly infinite) and $h_{rad}(x): = \Phi(|x|) \log |x|$
satisfies $\nabla h_{rad}\in L^\infty_{loc}$.
 Then all statements of
Theorem \ref{ACprec} for the Pauli operator with magnetic field $\mu$
are valid with $\Phi : = \Phi_{rad}+
\sfrac{1}{2\pi}\int \wt\mu^*(\rd x)$. 
\end{corollary}

{\it Proof.}
Let $\bA_{rad}:= \nabla^\perp h_{rad} \in L^\infty_{loc}$
and let $\wt h$ be the generating function of $\wt\mu$
given in Theorem \ref{thm:hdef} for some $\ep<\ep(\wt\mu)$.
Then $(\wt h, \bA_{rad})\in \cP^*$ with a magnetic field $\mu$,
hence the Pauli operators are well defined and unitarily equivalent.
Clearly $\pi^{\wt h,\bA_{rad}}(\psi,\psi) = \pi^h(\psi,\psi)$
with $h=h_{rad}+\wt h$, hence any  zero energy state $\psi$
must be in the form $\psi = (e^h g_+, e^{-h}g_-)$
where $g_\pm$ are (anti)holomorphic. 
Now we can follow the proof of Theorem \ref{ACprec}.
We use the estimates (\ref{A2}), (\ref{errorh}) for
the $\wt h$ part of the generating potential and
we estimate $h_{rad}(x): = \Phi(|x|) \log |x|$ 
by $\lim_{R\to\infty}\Phi(R)$ for large $R$, $\lim_{R\to0}|\Phi(R)|=0$
for small $R$ and by $\Phi \in L^\infty_{loc}(\bR)$ for intermediate $R$.
$\;\;\Box$

\section{A counterexample}\label{count}
 
In this section we present the construction of the
Counterexample \ref{cont}. For simplicity, the magnetic
field will  be only bounded and not continuous,
but it will be easy to see that a small mollification does
not modify the estimates.

Let $\delta < 1/10$ be a fixed small number
 and $N_k = 10k$ for $k=1, 2, \ldots$.
We denote the $N_k$-th roots of unity by $\zeta_{k,j}: = \exp(2\pi i j/N_k)$,
$j=1, 2, \ldots N_k$.
Let $D_{k,n,j}: =\{ x \; : \; |x-n\zeta_{k,j}|\leq\delta\}$
 be the disk of radius $\delta$ about $n\zeta_{k,j}$,
let  $\overline{D}_{k,n,j}: =\{ x \; : \; |x-n\zeta_{k,j}|\leq 2\delta\}$
be the twice bigger disk.

Let $0<\ep< 1/4$ be fixed. The magnetic field $B$  is given as
$B: = B_0 + \wt B - \wh B$
with
$$
	B_0(x) : = 2(1+\ep)\delta^{-2} {\bf 1}( |x|\leq \delta)
$$
$$
	\wt B := \sum_{k=1}^\infty \wt B_k, \qquad
	\wt B_k: =\sum_{n=4^k+1}^{4^k+2^k}\wt  B_{k,n},
	\qquad\wt  B_{k,n}(x): = \sum_{j=1}^{N_k} 2
	 \delta^{-2} {\bf 1}(x\in D_{k,n,j})
$$
and
$$
	 \wh B : =  \sum_{k=1}^\infty \wh  B_k, \qquad
	 \wh B_k: =\sum_{n=4^k+1}^{ 4^k+2^k} \wh  B_{k,n},
	\qquad  \wh B_{k,n}(x): = {1\over 2\pi}\int_0^{2\pi}
	\wt  B_{k,n}(|x|e^{i\theta})\rd \theta
$$

The field $\wt B$  consists
of uniform field "bumps" with flux $2\pi$ localized on the disks $D_{k,n,j}$
 around points $n\zeta_{k,j}$
that are located on  concentric circles of radius $n$.
The field $\wh B$  is 
 the radial average of $\wt B$.
The field $B_k=\wt B_k - \wh B_k$ is called the $k$-th {\it band}.
The relation (\ref{impr}) is straightforward by construction.

We define the  potential function
$h : = h_0 + \wt h - \wh h$, with
$$
	h_0(x):= {1\over 2\pi} \int_{\bR^2} \log|x-y|B_0(y)\rd y
$$
$$
	\wt h := \sum_{k=1}^\infty \wt h_k, \qquad
	\wt h_k: =\sum_{n=4^k+1}^{4^k+2^k}\wt  h_{k,n},
	\qquad\wt  h_{k,n}(x): =
	 {1\over 2\pi} \int_{\bR^2} \log|x-y|\wt B_{k,n}(y)\rd y -N_k \log n
$$
and
$$
 	\wh h := \sum_{k=1}^\infty \wh h_k, \qquad
	\wh h_k: =\sum_{n=4^k+1}^{ 4^k+2^k}\wh h_{k,n},
	\qquad\wh  h_{k,n}(x): =
	  {1\over 2\pi}  \int_0^{|x|} {\rd r\over  r} \int_{|y|\leq r}
	  \wh B_{k,n}(y)\rd y \; .
$$
Clearly $\Delta \wt h_k=\wt B_k$,  $\Delta \wh h_k=\wh B_k$.
Easy computations yield the following relations:
$$
	h_0(x) = (1+\ep) \log |x| \qquad \mbox{for} \quad |x|\ge \delta\; ,
$$
$$
	\wt  h_{k,n}(x) =  \sum_{j=1}^{N_k}\log \Big| \zeta_{k,j}-
	{x_1+ix_2\over n}\Big|
	=  \log\Big| 1- \Big( {x_1+ix_2\over n}\Big)^{N_k}\Big|
	\qquad \mbox{for} \quad x \not\in \bigcup_{j=1}^{N_k} D_{k,n,j}\; ,
$$
$$
	\wh h_{k,n}(x) = 
	\left\{ 
	\begin{array}{ccc}
	0 & \mbox{for} & |x|\leq n-\delta \cr
	N_k \Big[ \log {|x|\over n} + O(n^{-1}) \Big] & 
	 \mbox{for} & |x|\ge n-\delta \; .\cr
	\end{array}
	\right. \; 
$$

The infinite sums in the definition of $\wt h$ and
$\wh h$ are absolutely convergent, hence $h\in L^\infty_{loc}$.
The sum of  the $\wt h_{k}(x)$'s converges since
$$
	\sum_{k=1}^\infty \sum_{n=4^k+1}^{4^k+2^k} 
	 \Big| {x\over n}\Big|^{N_k} <\infty
$$
for each fixed $x$ and $\wh h_k(x)$ is actually zero for all
but finite $k$.
 Therefore we know that $\Delta h = B$
in distributional  sense.
Moreover,  we can rearrnge the sums and write
$$
	h = h_0 + \sum_{k=1}^\infty h_k ,
	\qquad  h_k: = \sum_{n=4^k+1}^{4^k+2^k} h_{k,n} \; ,
	\qquad  h_{k,n} := \wt h_{k,n} - \wh h_{k,n} \; .
$$

A short calculation shows that
for each $k_0$
\be
	\sum_{k=1\atop k\neq k_0}^\infty |h_k(x)| =O(1)
	\qquad \mbox{for} \quad 3\cdot 4^{k_0-1}-1 \leq |x|\leq 3\cdot 4^{k_0}
	+1 \; .
\label{out}\ee
Hence the size of $h(x)$ is determined by $h_0(x)$ and
the band nearest to $x$.

\bigskip

Now we show that $\int e^{-2h}=\infty$, in fact $\int_{D} e^{-2h}=\infty$
where $D: = \bigcup_{k,n,j} \Big( \overline{D}_{k,n,j}\setminus 
D_{k,n,j}\Big)$. The proof of $\int e^{2h}=\infty$ is similar but easier.
 We fix $k\ge 1$, $4^k + 1\leq m \leq 4^k+2^k$,
$1\leq \ell \leq N_k$ and let $x\in \overline{D}_{k,m,\ell}\setminus 
D_{k,m,\ell}$. Then
$$
	h_{k,n}(x) = \left\{ 
	\begin{array}{ccc}
	\log \Big| 1 - \Big({n\over x_1+ix_2}\Big)^{N_k}\Big| + N_k O(n^{-1})
	 & \mbox{for} & n<m \cr
	\log \Big| 1 - \Big({x_1+ix_2\over n}\Big)^{N_k}\Big|
	 & \mbox{for} & n>m \; .
	\end{array}
	\right. \; 
$$
Writing $x_1+ix_2= m\zeta_{k,\ell}+ (\varrho_1+i\varrho_2)$, 
$\varrho= (\varrho_1, \varrho_2)\in \bR^2$,  
$\delta \leq |\varrho|\leq 2\delta$
and expanding $h_{k,n}(x)$ around $m\zeta_{k,\ell}$ up
to second order in $\varrho$ we easily obtain that $h_{k,n}(x)\leq
N_k O(n^{-1})$ for each $n\neq m$ if $\delta$ is small enough, hence
 $h_k(x)\leq h_{k,m}(x) + O(1)$.

Moreover, $h_{k,m}(x) = \log \Big| 1- [(x_1+ix_2)/m]^{N_k}\Big| + O(1)$
since $|\wh h_{k,m}(x)|=O(1)$ for any $m-2\delta \leq |x|\leq m+2\delta$.
Hence
$$
	\int e^{-2h} \ge C \sum_{k=1}^\infty \sum_{m=4^k+1}^{ 4^k+2^k}
	{1\over m^{2(1+\ep)}}
	\sum_{\ell=1}^{N_k} \int_{ \overline{D}_{k,m,\ell}\setminus 
	D_{k,m,\ell}} \Big| 1 - \Big( {x\over m}\Big)^{N_k}\Big|^{-2}
	 \rd x
$$
$$
	= C \sum_{k=1}^\infty \sum_{m=4^k+1}^{ 4^k+2^k}
	{N_k \over m^{2(1+\ep)}}  \int_{\delta \leq |\varrho|\leq 2\delta}
	\Big| 1 - \Big( 1 + {\varrho_1+i\varrho_2\over m}\Big)^{N_k}\Big|^{-2}
	\rd \varrho
	\ge C 	 \sum_{k=1}^\infty {1\over N_k}\, 2^{k(1-4\ep)} =\infty\; .
$$

\medskip

Finally, we have to show that $e^{-h}{\bar f} \not\in L^2(\bR^2)$ for any
entire function $f$. 

First we show  that $f$ cannot have zeros. Suppose that
 $a$ is (one of) its zero closest to the origin, i.e.,
$f(z) = (z-a)^m g(z)$, $g$ is entire,
$g(0)\neq 0$, $m\ge 1$. Let $A_k:=\{ x \; : \;
3\cdot 4^k -1\leq |x|\leq 3\cdot 4^k +1\}$, then $h(x)=h_0(x) + O(1)$
for all $x\in A_k$ by (\ref{out}). Hence for a large enough $K$
$$
	\int e^{-2h}|f|^2 \ge C\sum_{k=K}^\infty
	\int_{A_k} e^{-2h_0(x)} |x-a|^{2m} |g(x)|^2 \rd x
$$
$$
	\ge C\sum_{k=K}^\infty 4^{2k(m-1-\ep)} \int_{A_k}|g(x)|^2 \rd x
	\ge  C|g(0)|^2\sum_{k=K}^\infty 4^{2k(m-1-\ep)}\cdot 4^k =\infty
$$
using that $|g|^2$ is subharmonic and the area of $A_k$ is 
of order $4^k$.

Now, since $f$ has no zeros, we can write $f=e^\varphi$ and we would
like to show that $\varphi$ is constant. It is enough to show that
$R:= \mbox{Re}\, \varphi$ is constant and we can assume $R(a)=0$. Suppose that
$\nabla R(a)\neq 0$ for some $a\in \bC$. Let $z_k$ be the point
where the maximum of  $R$ over the closed disk
$D_k:= \{ |x|\leq 3\cdot 4^k\}$ is attained. Since $R$
 is harmonic, $|z_k| = 3\cdot 4^k$.
Using (\ref{out}) and the  subharmonicity of $|e^{2\varphi}|$,  we have
\be
	\int e^{-2h}|f|^2 \ge C\sum_{k=1}^\infty 4^{-2k(1+\ep)}
	\int_{|x-z_k|\leq 1} |e^{2\varphi(x)}|\rd x
	\ge   C\sum_{k=1}^\infty 4^{-2k(1+\ep)} e^{2 R(z_k)}\; .
\label{pois}\ee
From the Poisson formula we easily obtain 
$|\nabla R(a)|\leq 4^{-k} \max_{D_k} R = 4^{-k} R(z_k)$
for large enough $k$. Hence $R(z_k) \ge 4^k|\nabla R(a)|$
and the integral in (\ref{pois}) is infinite. $\,\,\,\Box$.

\medskip

{\it Acknowledgements.} This work started during the first
author's visit at the Erwin Schr\"odinger Institute, Vienna.
Valuable discussions with T. Hoffmann-Ostenhof and M. Loss 
are gratefully acknowledged. The authors thank the referee
for careful reading and comments.

\newpage

\bigskip

AUTHORS' ADDRESS:

\bigskip
\bigskip

L\'aszl\'o ERD\H OS 

\vskip-10pt
 
School of Mathematics 
\vskip-10pt
Georgia Institute of Technology 
\vskip-10pt
Atlanta, GA 30332, USA 
\vskip-10pt
E-mail: {\tt lerdos@math.gatech.edu}

\bigskip

Vitali VOUGALTER  
\vskip-10pt
Department of Mathematics
\vskip-10pt
University of British Columbia
\vskip-10pt
Vancouver, B.C. Canada V6T 1Z2
\vskip-10pt
E-mail: {\tt vitali@math.ubc.ca}

\end{document}